\documentclass[]{spie}  

\usepackage{amsmath,amsfonts,amssymb}
\usepackage{graphicx}
\usepackage[table]{xcolor}
\usepackage[colorlinks=true, allcolors=blue]{hyperref}
\usepackage{comment}
\usepackage[gen]{eurosym}
\usepackage{siunitx}
\usepackage{pgfplots}
\pgfplotsset{compat=1.17} 
\usepackage{soul}
\usepackage{tikz}
\usepackage{tabularx}
\usepackage{multirow} 
\usepackage{graphicx}
\usetikzlibrary{positioning,arrows.meta}

\title{Additive manufacturing applications in astronomy: a review}

\author[a]{Younes Chahid}
\author[a]{Carolyn Atkins}
\author[a]{Greg Lister}
\author[a]{Rhys Tuck}
\author[a]{Stephen Watson}
\author[a]{Katherine Morris}
\author[a]{David Isherwood}
\author[a]{Jonathan Strachan}
\author[a]{Joel Harman}
\author[b]{Pearachad Chartsiriwattana}
\author[c]{Deno Stelter}
\author[d]{Werner Laun}

\affil[a]{UK Astronomy and Technology Centre, Royal Observatory, Edinburgh, EH9 3HJ, UK}
\affil[b]{National Astronomical Research Institute of Thailand, Chiang Mai, Thailand}
\affil[c]{University of California Observatories, California, United States}
\affil[d]{Max-Planck Institute for Astronomy, Heidelberg, Germany}

\authorinfo{Further author information:\\ E-mail: younes.chahid@stfc.ac.uk\\}

\begin{document} 
\maketitle

\begin{abstract}

Despite the established role of additive manufacturing (AM) in aerospace and medical fields, its adoption in astronomy remains low. Encouraging AM integration in a risk-averse community necessitates documentation and dissemination of previous case studies. The objective of this study is to create the first review of AM in astronomy hardware, answering: where is AM currently being used in astronomy, what is the status of its adoption, and what challenges are preventing its widespread use? The review starts with an introduction to astronomical instruments size/cost challenges, alongside the role of manufacturing innovation. This is followed by highlighting the benefits/challenges of AM and used materials/processes in both space-based and ground-based applications. The review case studies include mirrors, optomechanical structures, compliant mechanisms, brackets and tooling applications that are either in research phase or are implemented.
\end{abstract}


\section{INTRODUCTION}
\label{sec:intro}  

Astronomical instruments are used to convert the light (photons) collected from ground-based or space-based telescopes into data that describe our universe. The ultimate driving factor when designing a telescope is the number of photons that can be collected, which often results in larger structures with greater mass, a higher number of components, and increased complexity, making them more challenging and expensive to build. 

\begin{figure}
    \centering
    \includegraphics[width=0.8\linewidth]{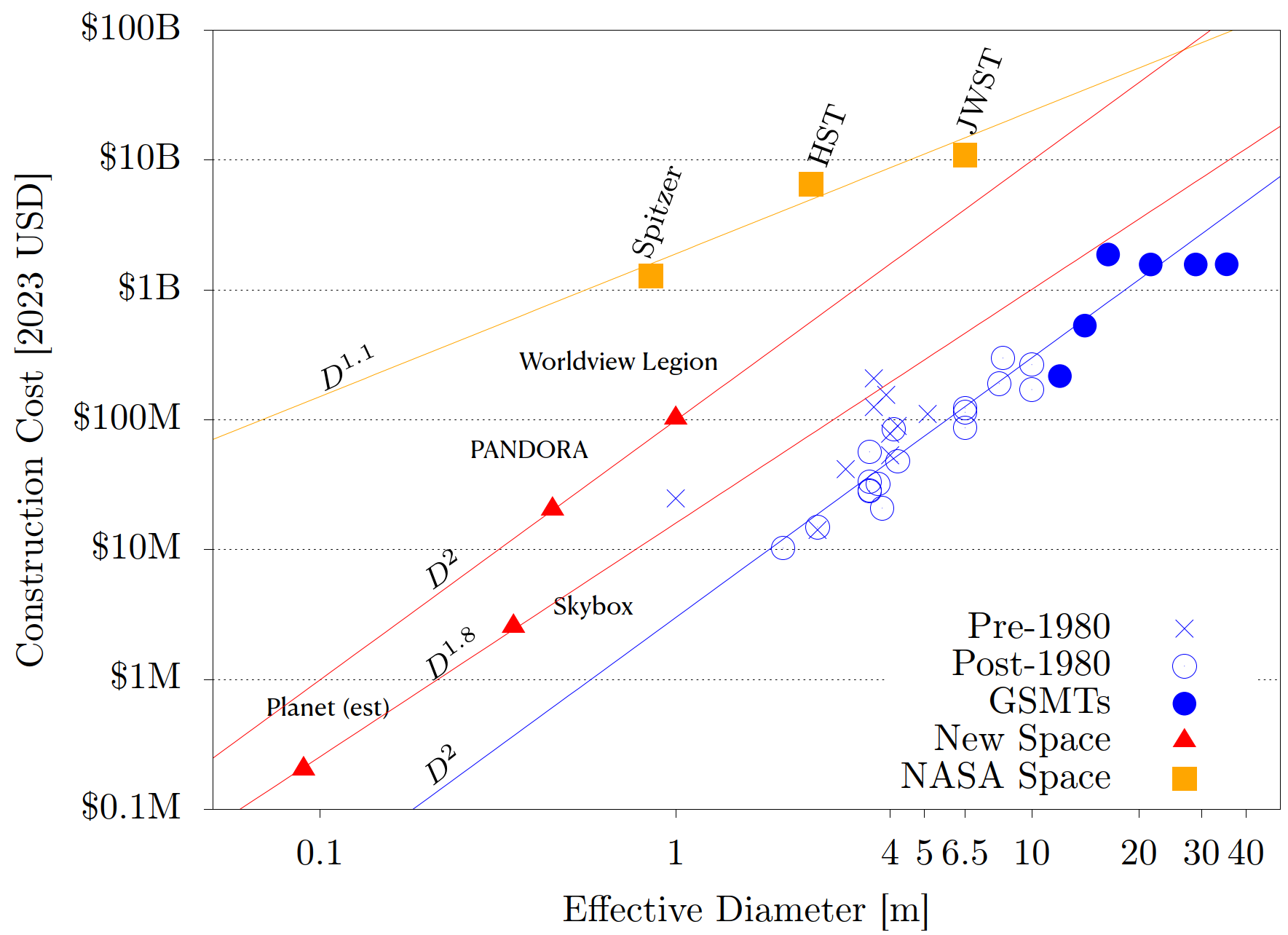}
    \caption{Relationship between construction cost and effective diameter of space-based and ground-based telescopes \cite{douglas2023approaches}. This graph shows the cost halving between Pre-1980 and Post-1980, caused by the switch to honeycomb and segmented mirror technologies in ground-based observatories. GSMTs refer to Giant Segmented Mirror Telescopes. Image credit: Douglas, et al,. (2023)~\cite{douglas2023approaches}.}
\label{fig:cost_telescopes}
\end{figure}

For example, when it comes to ground-based telescopes, the primary mirror diameter has increased by over 100 times in the past three centuries \cite{graves2019precision}. This increase in size tends to correlate with significant increase in cost. For instance, the Keck observatory and Gran Telescopio Canarias (GTC) telescope have an effective diameter of $\sim10$ meters and a construction cost of $\sim\$100$ million \cite{van2004scaling}, compared to the Extremely Large Telescope (ELT) which has an effective diameter of $\sim40$ meters and an estimated cost of $\sim\euro1.45$ billions \cite{FAQELTES12:online}. This size to cost increase trend is present in both ground-based and space-based telescopes, as seen in Figure \ref{fig:cost_telescopes}. To better illustrate the magnitude of these costs, a comparison can be made between the cost-to-mass ratio of the International Space Station (ISS) and the James Webb Space Telescope (JWST). The ISS has a cost-to-mass ratio of \$337,000 per kilogram, while the JWST has a cost-to-mass ratio of \$1,700,000 per kilogram \cite{mcclelland2022generative}. This five folds price difference highlights the demanding field of astronomy particularly in developing highly sensitive instruments operating in harsh environments like space.

\subsection{Importance of manufacturing innovation}

Cost reduction is critical in developing and maintaining astronomical instruments as it directly relates to the significant challenge posed by the scarcity of funding sources in this field. Maturing enabling technologies in manufacturing has played a significant role in reducing costs. Innovations like honeycomb and segmented mirror technologies in ground-based observatories have played a substantial role in halving the cost of ground-based telescopes in pre/post 1980\cite{douglas2023approaches}. This evolution has had a lasting impact, with all of optical telescopes over 10 meters now employing segmented mirrors \cite{graves2019precision}.

Additive manufacturing (AM) has the potential to be the next important enabling technology, and has demonstrated a steady rise in market size\cite{hubsPrintingTrend}, sales of AM systems\cite{wohlersassociatesWohlersReport}, and increasing adoption across various industries. For example, in the medical industry, hip implant acetabular cups are increasingly additively manufactured with more than 150,000 made by LimaCorporate \cite{siemensLimaCorporateCreates}. In the aviation industry, Boeing products include 70,000 additively manufactured products \cite{boeing}. While in the space sector, Thales Alenia Space has 350 polymer parts and 79 metal components that are additively manufactured and currently in orbit \cite{3dnativesThalesAlenia}. 

\begin{figure}
    \centering
    \includegraphics[width=0.8\linewidth]{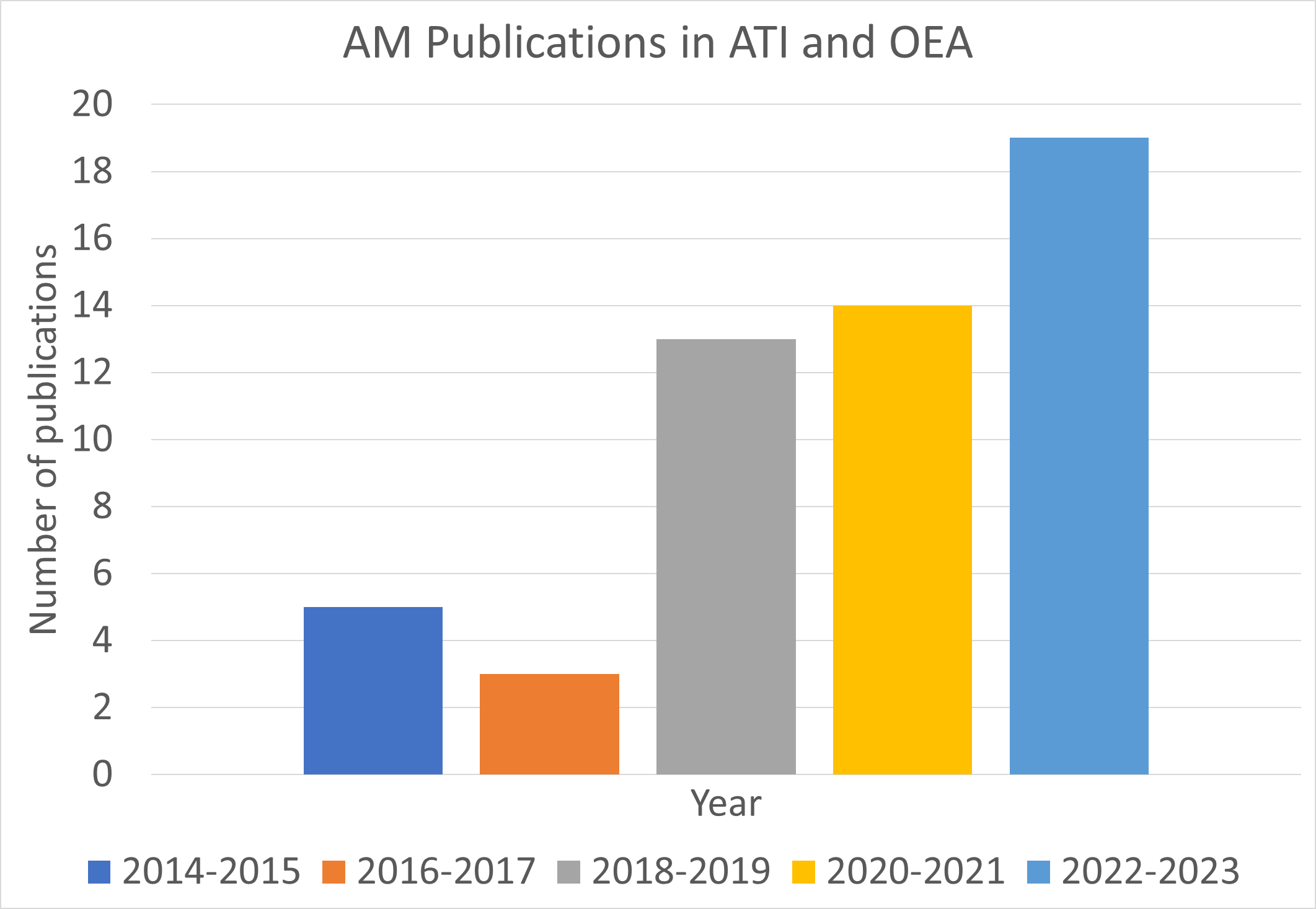}
    \caption{Trend of AM publications within ATI and OEA.}
    \label{fig:publications}
\end{figure}

Within the field of astronomy, there has been an increase in dedicated AM research. This growth can be quantified by looking at publications with AM in the title, from two field specific conferences: Astronomical Telescopes + Instrumentation (ATI) and Optical Engineering + Applications (OEA). The search criteria included other variants of the word AM like 3D printing, 3D-printed, additively manufactured etc. This growing trend, illustrated in Figure \ref{fig:publications}, reflects the increasing interest of AM use within the astronomical instrumentation community.

However, despite AM being identified by the European Space Agency (ESA) and the National Aeronautics and Space Administration (NASA) as an emerging technology \cite{esaFutureHere}, the rate of use of AM in astronomical applications remains limited and scarcely reported when compared to other sectors \cite{schnetler2020h2020}. Previous AM literature in this field include ``A2IM: Cookbook An introduction to additive manufacture for astronomy" which disseminated multiple types of case studies, materials and AM processes, with a broader focus on astronomical mirror developments. Other AM reviews in the adjacent field of optics are also found in literature \cite{zhang2021design,zolfaghari2019additive,jia2021research}. The aim of this review is to address a gap in existing literature by examining the broader field of AM applications in astronomical instruments. The following sections will include the review methodology, detailing inclusion/exclusion criteria. Subsequent sections will then explore AM use in different applications, highlighting its benefits and challenges. 

\section{AM applications in astronomy}

\subsection{Review Methodology}

In this review, a total of 102 AM parts/designs have been compiled and sourced from various platforms such as SPIE Digital Library, academic journals, news media, and manually collected data.  The review relies heavily on AM case studies shared in the public domain. However, the manual collection process has uncovered previously undocumented AM applications, like ones used in instruments parts GTC, Very Large Telescope (VLT) and ELT. This suggests that AM is potentially present in more projects but is just not shared in public domain. The lack of publications covering AM applications can be considered one of the challenges of AM adoption.

When collecting case studies, priority was given to those that reached the AM stage and produced a physical part. However, research focused solely on the design stage was also included. Case studies were divided into five distinct categories, each with its own subcategories, as outlined below:

\begin{itemize}
    \item \textbf{Environment}: This category highlights if the case study is intended for a space-based application or ground-based application. This is usually mentioned in the publication or deducted from the context of the research project. Space-based applications include astronomy and planetary science projects (for example, interplanetary spacecrafts and asteroid orbiters) as well as Earth observation projects. This section excludes commercial satellites.
    \item \textbf{Readiness}: This category is split between researched applications and implemented applications. Research stage means that the part was a demonstrator and has not been implemented in the final intended application. Implemented components include those deployed in the final version of the assembly or those that have passed or are still in the testing phase but are intended for use in the final version of the studied components, rather than serving as demonstrators.
    \item \textbf{Type of component}: A total of five types have been collected ranging from mirrors, optomechanical structures, compliant mechanisms,  brackets and tooling. To narrow the focus of the review paper, excluded AM applications include optical lenses \cite{heinrich2016additive}, waveguides \cite{kilian2017waveguide,chu2022additive} and antennas \cite{aziz2024characteristics}. 
    \item \textbf{Material}: Materials were classified into three main cateogries, metals, ceramics and polymers. Type of alloy or subcategory name of the material is also documented.
    \item \textbf{AM process}: This category was split into seven processes: powder bed fusion (PBF), directed energy deposition (DED), material extrusion (ME), material jetting (MJ), binder jetting (BJ), sheet lamination (SL), and vat photopolymerization (VPP). These are the same categories listed in ISO/ASTM 52900:2021\cite{isoISOASTM529002021} which defines AM terminology. Different terminology is sometimes used in literature like Selective Laser Melting (SLM) or Direct Metal Laser Sintering (DMLS) which falls in the PBF category, specifically Laser PBF (LPBF,L-PBF). Another subcategory of PBF is Electron 
    Beam Melting (EBM) which uses an electron beam instead of a laser. Other terminology include Fused Deposition Modeling (FDM) or Fused Filament Fabrication (FFF) of which both are part of material extrusion while Stereolithography (SLA) is part of the VPP category. This review will use these terms interchangeably.
\end{itemize}

\subsection{Review Result}

\begin{figure}
    \centering
    \includegraphics[width=0.6\linewidth]{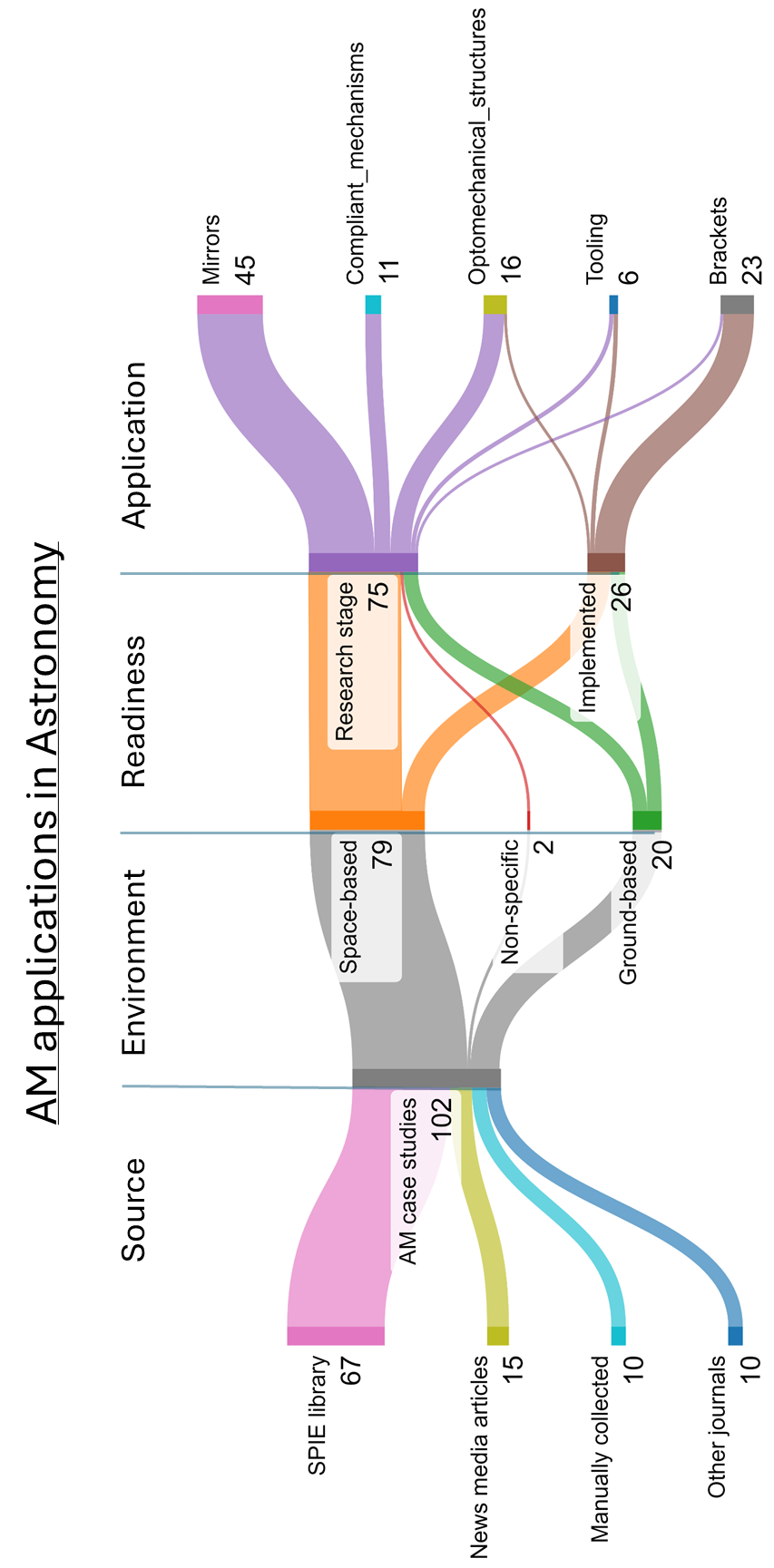}
    \caption{Diagram showing the different AM applications in astronomy starting from the source to environment, readiness and type of application.}
    \label{fig:sankey1}
\end{figure}

A summary of the review results, seen in Figure \ref{fig:sankey1} (based on data compiled from Table \ref{tab:Compilation1} \& \ref{tab:Compilation2}), shows all compiled types of applications categorised to the different sources of publications, use environment and readiness level. While not exhaustive of all literature in this field, this summary can serve as a guide to understanding current research trends/gaps. The following sections will discuss the benefits and challenges of AM in each application.

\subsubsection{Mirrors}

\begin{figure}
    \centering
    \includegraphics[width=1\linewidth]{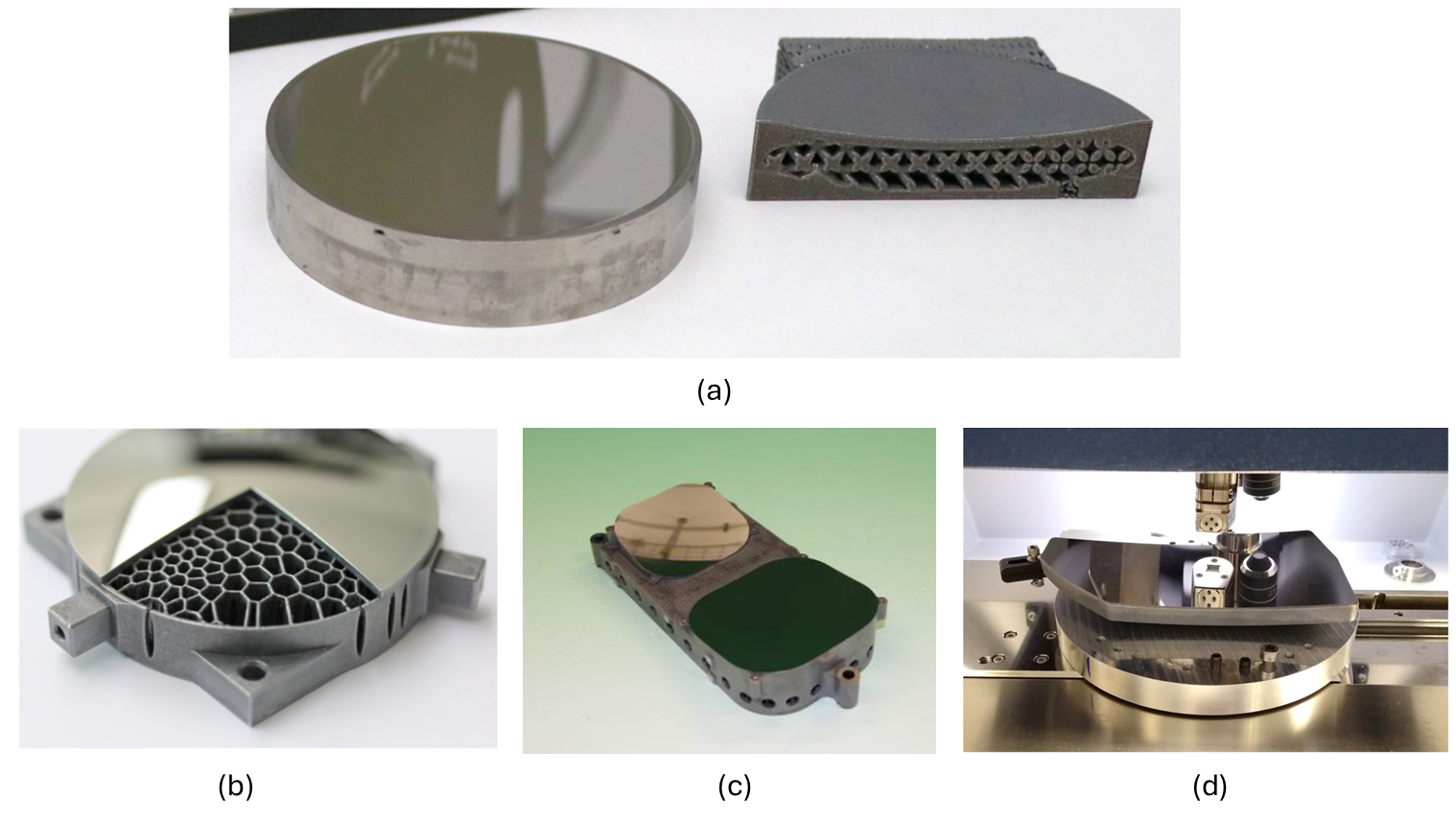}
    \caption{Examples of additively manufactured lightweight mirrors (a\cite{snell2020additive},b\cite{hilpert2018precision},c\cite{heidler2018additive},d\cite{westsik2023design}). Image credits: (a) Snell, et al,. (2020)~\cite{snell2020additive}; (b) \textcopyright \, Fraunhofer-IOF  (2018)~\cite{hilpert2018precision};(c) \textcopyright \, Fraunhofer-IOF (2018)~\cite{heidler2018additive};(d) Westsik, et al,. (2023)~\cite{westsik2023design}}
    \label{fig:mirrors2}
\end{figure}

Telescope mirrors are crucial components in astronomical instruments, as they directly impact the quality and accuracy of the observations. The required optical quality of a mirror is dependant on the operating wavelength \cite{a2imcookbook}. For example, and while it is challenging to reduce surface specifications to single values, mirrors designed for visible light wavelength require a surface roughness of $<$5 $nm$ root mean square (RMS), while surface roughness of mirrors operating in longer wavelength like infrared can be $<$12 $nm$ RMS \cite{atkins2019additively}. Another important factor in mirror designs is their mass. Lightweight mirrors are essential in reducing structural demands of telescopes, enabling larger apertures and more stable mounts. This reduction in mass facilitates the construction of larger telescopes both on ground and especially in space, allowing for lower launch costs, portability and ease of deployment. Examples of additively manufactured lightweight mirrors can be found in Figure \ref{fig:mirrors2}.

\begin{figure}
    \centering
    \includegraphics[width=.6\linewidth]{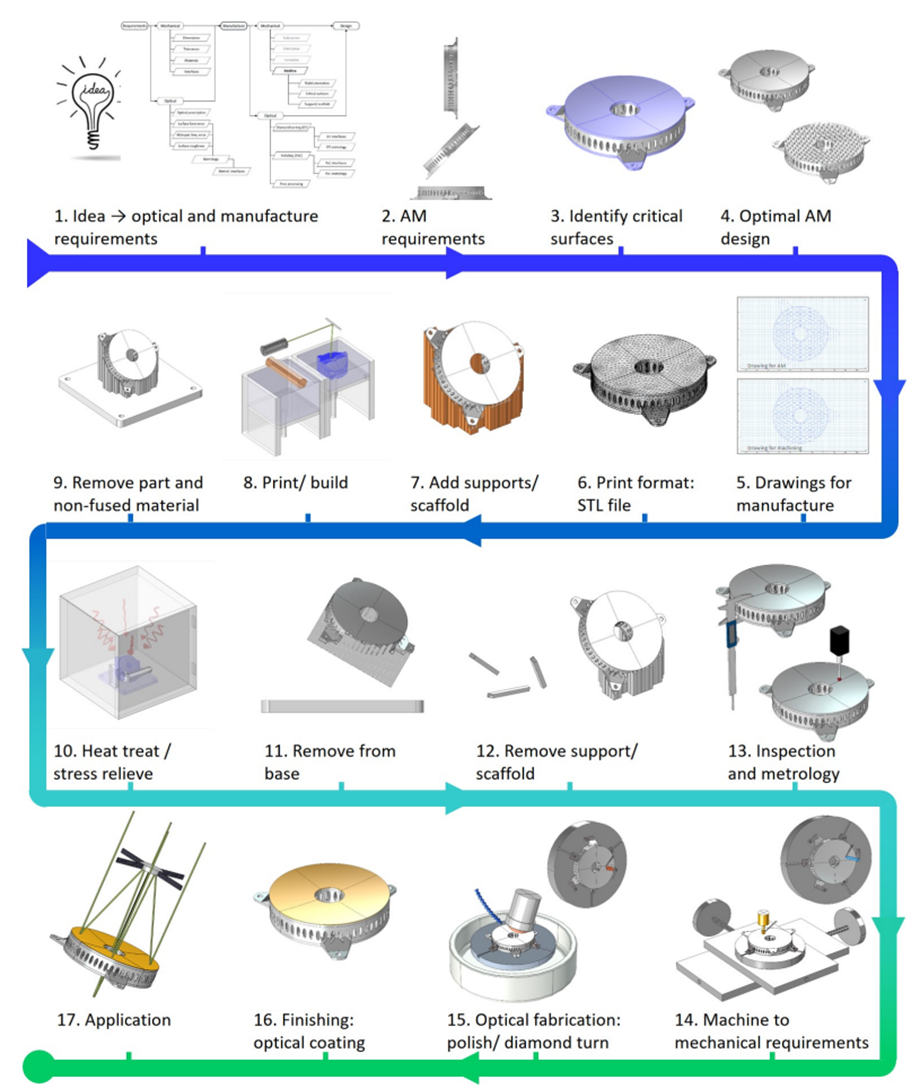}
    \caption{Typical process chain of producing AM mirrors\cite{atkins2022opticon}. Image credits: Atkins, et al,. (2022)~\cite{atkins2022opticon}}
    \label{fig:mirror_process}
\end{figure}

Conventional lightweight mirror designs can be classified into three types: contoured back, open-back, and sandwich \cite{a2imcookbook}. Sandwich structures offer a good combination in terms of rigidity and lightweighting, but they are conventionally more expensive and time consuming to make compared to contoured back or open-back structure designs, which are cheaper and faster to make but offer less rigidity. A rigid mirror substrate ensures that the mirror maintains its shape and alignment under conditions associated with operation and the manufacturing process chain (Figure \ref{fig:mirror_process}). 

Compiled research on AM mirrors (Table \ref{tab:Compilation1}) is predominantly targets space environments, a preference attributed to the elevated emphasis on lightweighting in space applications. Another likely explanation is the size limitation of current AM processes, which is not mature enough to compete with processes optimised for larger ground-based mirrors. In terms of material choice, the prevailing material was aluminum, due to its competitive specific stiffness of the material (ratio of stiffness to density). Aluminum provides numerous benefits, including affordability, non-toxicity, and ease of machining. This material simplifies not only the fabrication process, but also structural integration, since it matches the coefficient of thermal expansion (CTE) of  optomechanical structures, which are also usually in aluminium. Other material choices for mirrors include fused silica glass, glass ceramic (Zerodur), silicon carbide (SiC), and ultra-low expansion (ULE) glass. These materials offer advantages such as superior thermal stability, lower thermal expansion, and higher stiffness. A comparison of these materials is further discussed in literature\cite{garoli2020mirrors}.

Compared to other AM applications, a unique challenge in producing AM mirrors is the higher sensitivity to porosity, which can affect both the structural integrity of the mirror and the surface finish quality. AM porosity can be caused by multiple factors related to raw material or process settings \cite{yang2022quality}. Porous mirrors would often require coating in Ni or nickel-phosphorus (NiP), removing any porosity concern \cite{atkins2017additive}. Coating deposits a metal layer than can be DT or polished. For example, a \SI{100} {\micro\meter} NiP coating was added to an AlSi10Mg AM mirror, prior to DT process \cite{atkins2019additively}. AM of AlSi40 has also been researched, due to its closer CTE to NiP coating. For example, an NiP layer of approximately \SI{55} {\micro\meter} was applied on all outer and interior surfaces of an AM mirror with a lightweight voronoi type lattice, allowing for 63.5\% mass reduction. Post-processing using DT and polishing led to $\sim$\SI{1} {\nano\meter} RMS \cite{hilpert2018precision, hilpert2019design,eberle2019additive}.

However, coatings introduce an additional process step that is limited to the material CTE compatibility. An alternative to tackle AM porosity is to use hot isostatic pressing (HIP) to densify the component. This step works by subjecting the AM mirror substrate to inert gas, high temperatures and high pressures causing the closure of pores \cite{ge2023post}. Previous research showed decreased porosity of AM mirror blanks after using HIP \cite{tan2020design}. The remaining porosity is likely due to the pores or cracks being connected to the external surface, which cannot be closed by HIP \cite{tammas2016effectiveness,herzog2015optical,snell2022towards}. Also, HIP decreases the material ultimate tensile strength (UTS) and increases ductility, making the material less brittle and prone to distortion, suggesting a follow up heat treatment \cite{eberle2019additive}, which increases the fabrication process chain.

\subsubsection{Optomechanical structures}

Optomechanical structures serve to combine optical components such as lenses, mirrors, and prisms with mechanical elements like cells, housings, and trusses to create an optical instrument \cite{SPIE}. The key requirement is to preserve the precise alignment of these optical components under different operational conditions. This necessitates minimising structural deflections due to gravity, temperature fluctuations and vibrations/shocks during transport or use, in order to avoid misalignment or optical distortions. Choosing the appropriate material is critical to ensure thermal stability, durability, and compatibility with optical components. For example, the chosen material of an optomechanical structure should ideally have a similar CTE to the optical component they are paired with, to minimise differential expansion or contraction. Commonly used materials for optomechanical structures include metals like aluminium, stainless steel and titanium \cite{SPIE}.

\begin{figure}
    \centering
    \includegraphics[width=1\linewidth]{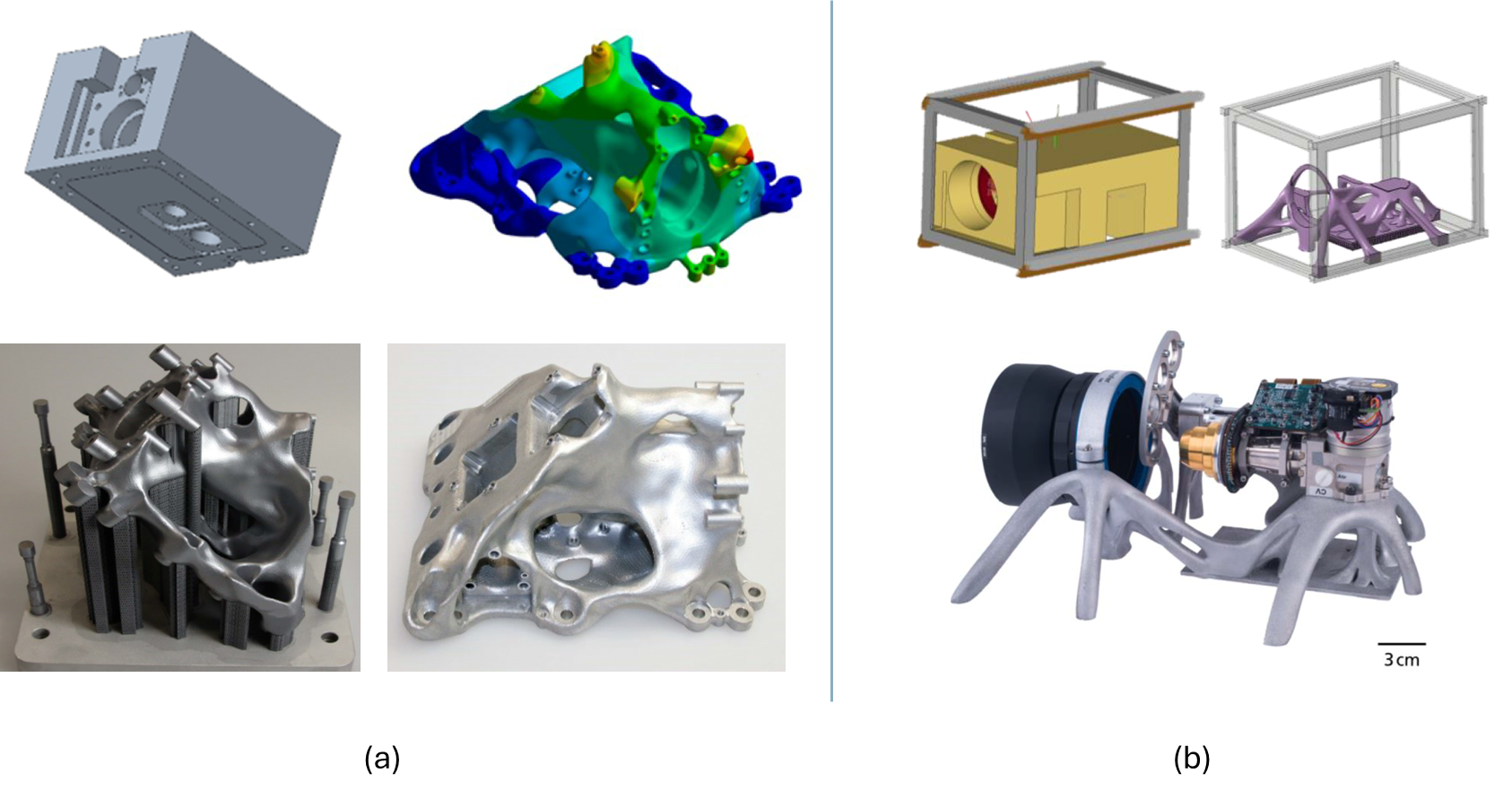}
    \caption{Redesign of a space-based optical housing produced using LPBF in AlSi40\cite{heidler2019topology} (a). Redesign of a 12U Earth observation Cubesat optical bench, produced using LPBF in Scalmalloy \cite{schimmerohn2018additive} (b). Image credits: (a) \textcopyright \, Fraunhofer IOF~\cite{heidler2019topology}; (b) Schimmerohn, et al,. (2020)~\cite{schimmerohn2018additive}}
    \label{fig:opto_space1}
\end{figure}

Multiple studies (Table \ref{tab:Compilation2}) show AM benefits when applied on optomechanical structures. For instance, AM was utilised to redesign a space-based optical housing \cite{heidler2019topology}, resulting in a 20\% mass saving while meeting the specified thermal and mechanical specifications. Implementing AM helped replace the conventional beryllium and aluminum metal matrix (AlBeMet), also known for associated production risks. To achieve this, the chosen AM alternative material was AlSi40, which consists of 40\% mass silicon, exhibiting good mechanical properties and a CTE close to AlBeMet. This substitution helped avoid the need for a thermo-elastic redesign of the entire optomechanical system. The redesign employed topology optimisation and finite element method (FEM) simulations, to meet different vibration and thermal boundary conditions. AM was done using LPBF and included multiple tensile specimens to facilitate material testing (Figure \ref{fig:opto_space1} a). Typical metal LPBF surface roughness values, expressed as the arithmetical mean deviation of the profile (Ra), tend to exceed $\sim$\SI{20}{\micro\meter} \cite{li2018research}, which meant that a 5-axis milling machine was used to remove excess material and incorporate tapped holes with precise tolerances. An X-ray Computed Tomography (XCT) scan of \SI{120}{\micro\meter} voxel size did not detect defects or pores and the tensile tests revealed properties comparable to the bulk material ones. A previous study \cite{mueller2019microstructural} shows how Al40Si not only met the required CTE of the bulk material AlSi40 but also surpasses its mechanical properties by nearly 20\% and while having a relative density of 99.5\%.   

A subcategory of optomechanical structures that can take advantage of AM are optical benches, which serve as specialised platforms designed to precisely support and align optical components. For example, an optical bench of a 12U Earth observation Cubesat was redesigned using AM to allow for improved heat transfer capabilities and lightweighting, while conforming to vibrational requirements \cite{schimmerohn2018additive}. The first benefit stemmed from consolidating heat transfer and emission functionalities within a single, compact design. This was achieved by using a 3D pyramid geometry, allowing for a net power emission increase of 36.5\% without the use of a dedicated heat transfer device. The second benefit was lightweighting, which was achieved by defining a stiffness goal (from acceleration and vibration loads) within a topology optimisation study. The result was a lightweighted model (Figure \ref{fig:opto_space1} b) with a significantly lower stress to yield value. The final design was 3D printed using LPBF process and Scalmalloy\textsuperscript{\textregistered}, which is a high-performance aluminum-magnesium-scandium alloy, with superior tensile and fatigue properties compared to AlSi10Mg \cite{awd2017comparison, muhammad2021comparative}. Post processing steps included sandblasting, machining, grinding and tapping at component interfaces. Part verification included vibrational testing, which was completed using a shaker table and of which the results conformed to simulation results.

The readiness of optomechanical applications is increasingly evident, as shown in a study\cite{esaPrintingTelescope} conducted by ESA for NASA's Earth Observation System (EOS) Aura space-based mission. While the final mission used conventional manufacturing methods, an AM alternative ``demonstrator telescope" was developed to assess AM general readiness. The use of the topology optimisation method in combination with AM in aluminium, led to a mass reduction of 73\% (from 2.8kg to 0.76kg), with no reduction in measurement quality. The AM equivalent was consolidated into fewer components and had four attachment points instead of six attachment points. AM part validation included successful thermal, vibration and shock experiments. These tests made the AM ``demonstrator telescope" optomechanical structure, compliant with the optical performance requirements\cite{esaPrintingTelescope}. Still in the relatively small/medium scale, another AM adopted study include an optomechanical structure 3D printed in polymer and operating from the ISS, as part of the cosmic Ultraviolet (UV) telescope. The study used Ultem 9085 polymer, which is a polyetherimide (PEI) thermoplastic with high strength to mass ratio and low outgassing \cite{padtincPrintingSpace}. The use of AM helped the research team in meeting both ISS safety and mass restrictions while providing the necessary structural support needed \cite{nessancleary3DprintedTelescope}.

\begin{figure}
    \centering
    \includegraphics[width=1\linewidth]{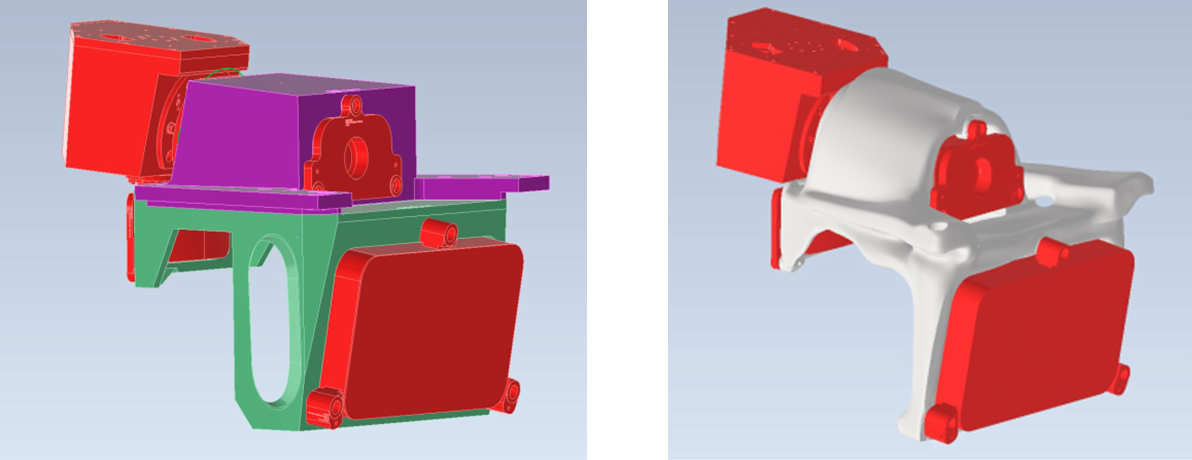}
    \caption{Study investigating use of AM for the TMA structure ELT's METIS instrument using DED process in aluminium 5356\cite{wells2023lightweighting}. Image credits: Wells, et al,. (2023)~\cite{wells2023lightweighting}}
    \label{fig:opto4}
\end{figure}

In a similar way to AM mirrors, most researched optomechanical structures have a relatively small size, which can again be explained by the limited build size of the commonly used AM machines. However, optomechanical structures of considerable size can also gain advantages from AM, as illustrated by projects like ESA's Advanced Telescope for High Energy Astrophysics (ATHENA), a space-based X-ray observatory. In this study\cite{schneider2023additive}, the optical bench has a diameter of 2.5 meters, is 3D printed in Ti6Al4V using DED process, and is designed to contain 600 mirror module pockets. The use of AM enabled the elimination of the high expenses linked with acquiring large titanium forgings and extensive machining. A further example of DED in a ground-based telescope instrument is from a research\cite{wells2023lightweighting} investigating the structure redesign of a three mirror anastigmat (TMA) for mid-infrared ELT imager and spectrograph (METIS). The study's optomechanical redesign demonstrated a 30\% reduction in mass and consolidation of 30 parts into a single component. FEM simulation met the required optical, structural, and modal criteria. The investigated material was aluminium 5356 and due to the relatively larger size of 615 mm × 530 mm × 525 mm (see Figure \ref{fig:opto4}). A detailed exploration of the state of DED AM process can be found in literature\cite{lehmann2022large, costello2023state}. 

\subsubsection{Compliant mechanisms}

Compliant mechanisms (CM) are crucial in high-precision or long-lifetime applications, the case of many astronomical instruments, offering distinct advantages over classical mechanisms where friction-related issues such as the need for lubrication, debris generation, backlash, and stick-slip can be problematic\cite{kiener2022validation}. To address these challenges, CM can achieve different motions like linear or rotary ones within the range of few millimeters with minimal issues, showcasing exceptional fatigue behavior, thanks to the elastic behavior of these structures\cite{kiener2022validation}. However, the intricate nature of compliant mechanisms has traditionally demanded sophisticated and costly manufacturing methods, such as wire Electro-Discharge Machining (EDM), involving substantial material waste, meticulous assembly procedures, and in some cases unwanted brittleness in heat affected zones \cite{morris2022additively}. The advent of AM offers a unique opportunity to streamline the production of compliant mechanisms, reducing costs, minimising material waste, and consolidating intricate designs with unparalleled precision, thus allowing their integration into astronomical instruments.

\begin{figure}
    \centering
    \includegraphics[width=0.75\linewidth]{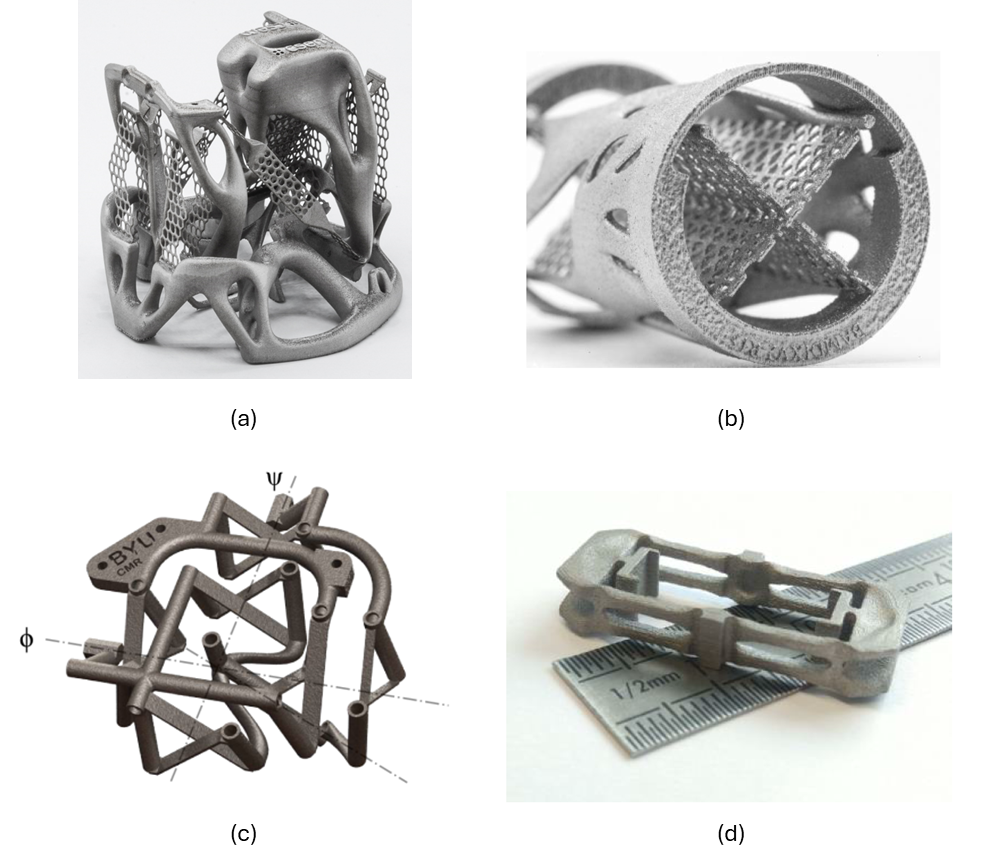}
    \caption{rendering of CRTM (a) and C-flex type pivot (b), both printed in 17-4PH Stainless Steel using PBF (Laser) \cite{kiener2022validation}. 2 DOF space pointing mechanism 3D printed in titanium using PBF (beam) \cite{merriam2013monolithic}. Actuator driven flexure 3D printed in 316L stainless steel using binder jetting AM process \cite{atkins2022opticon, morris2022additively}. Image credits: (a) CSEM; (b) Kiener, et al,. (2022)~\cite{kiener2022validation}; (c) Merriam, et al,. (2013)~\cite{merriam2013monolithic};(d) Atkins, et al,. (2022)~\cite{atkins2022opticon}}
    \label{fig:compliant1}
\end{figure}

Studies from literature include a Compliant Roto-Translation Mechanism (CRTM) that has an two directional rotational motion of \SI{\pm 15}{\degree} and linear motion of ±5 mm suitable for combination with actuators implemented in instrument scanners or interferometers. The component, seen in Figure \ref{fig:compliant1} (a), was additively manufactured in 17-4PH stainless steel using LPBF and designed using topology optimisation with multiple constraints related vibration, symmetry, maximum displacement/stresses and also overhang angles \cite{kiener2022validation}. AM Part validation followed ESA's ECSS-Q-ST-70-80C \cite{ecssECSSQST7080Cx2013} standard, which requires including witness samples on the build plate. These samples underwent density, metallography, and tensile tests, along with bending fatigue tests tailored to this specific application. 

A second CM example is a ``C-flex'' pivot flexure with crossed blades, seen in Figure \ref{fig:compliant1} (b). These pivots are integrated in a mirror tile (tip-tilt mechanism), and their objective is to adjust the position of a mirror, correcting for assembly errors. This flexure enabled manipulation of the mirror in three degrees of freedom. Furthermore, several measurements showed positive results of overall mirror position accuracy surpassing project requirements. Additionally, the use of AM allowed for part consolidation and to the AM of a monolithic flexible pivot components within few days instead of the conventionally time consuming machining, aligning, and welding in which any small design change would require different tooling\cite{kiener2022validation}. The chosen material and AM process were again 17-4PH stainless steel and LPBF. The selection of 17-4PH stainless steel is motivated by the need for a fatigue limit sufficient to counter the extensive number of cycles of the product \cite{kiener2022validation}. 

Further examples include a two degrees of freedom space pointing mechanism printed in titanium using EBM\cite{merriam2013monolithic} (Figure \ref{fig:compliant1} (c)), and a monolithic flexure which holds a linear actuator to deliver a displacement to a mirror surface. The monolithic flexure was printed in AlSi10Mg, 316 Stainless Steel and Ti6Al4V via two AM processes (PBF and binder jetting), and exhibited a 50\% mass reduction through the use of topology optimisation.

\subsubsection{Brackets}

Brackets are structural or mechanical components designed to support, reinforce, or attach various parts of a system or structure. They serve as connecting or mounting points for other components and can be of various shapes and sizes, depending on their specific application and the loads they are intended to bear.

\begin{figure}
    \centering
    \includegraphics[width=.6\linewidth]{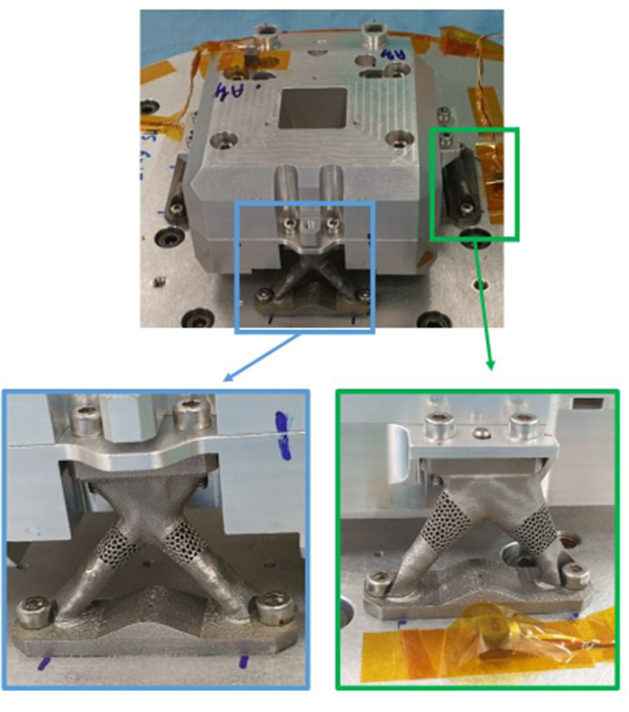}
    \caption{Redesign of a satellite support bracket 3D printed using LPBF in titanium Ti6AL4V\cite{manil2021structural}. Image credits: Manil, et al,. (2021)~\cite{manil2021structural}}
    \label{fig:bracket_space2}
\end{figure}

For example, a study from literature shows the application of AM in the redesign of a satellite support bracket for the Space-based multi-band astronomical Variable Objects Monitor (SVOM) \cite{manil2021structural}. These support brackets are designed to bear the mass of the camera, and serve as the connecting mount to the satellite chassis while maintaining alignment. The design process explored lattice structures and topology optimisation variants for three titanium Ti6Al4V delta shaped support feet, each weighing 26g and with a bounding box 40x20x15mm (see Figure \ref{fig:bracket_space2}). The AM designs successfully met three main objectives: first, to provide thermal insulation of the detector (\SI{-35}{\celsius}) from the lower plate (\SI{0}{\celsius}). Second, to withstand the dynamic forces and vibrations experienced during launch, ensuring the components remain secure and functional, and third, to be lightweight.

\begin{figure}
    \centering
    \includegraphics[width=0.8\linewidth]{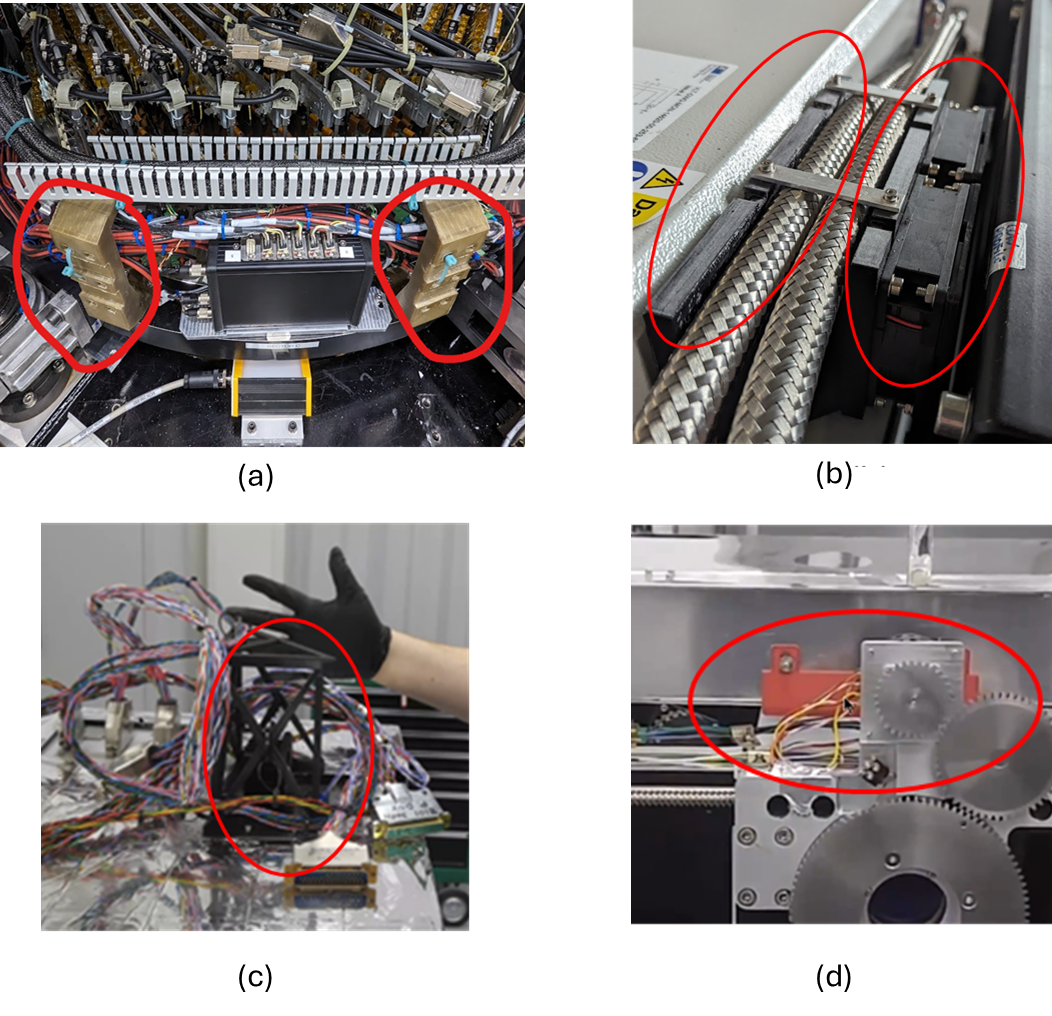}
    \caption{VLT MOONS instrument Antero cable support (a) and ASA air duct (b) done by Watson S. and Strachan J. respectively using FDM AM process. Cable holder (c) and cable protector (d), both 3D printed in ABS using FDM process and both used for GTC CIRCE instrument and done by Stelter D. Image credits: (a) Watson S. ; (b) Strachan J.;(c,d) Stelter D.}
    \label{fig:groundbraqcket}
\end{figure}

Further space-based studies include brackets on JUpiter ICy moons Explorer (JUICE) mission by ESA and the Psyche mission by NASA. JUICE is an interplanetary spacecraft that launched in April 2023 and is expected to reach Jupiter in 2031 \cite{voxelmattersJUICEPrinted,esaJuiceFactsheet}. The spacecraft included 13 metal brackets \cite{esaScienceamp} made in Scalmalloy material and produced using LPBF process. These brackets, play the role of the interface between the primary structure and the scientific payload and equipment like reaction wheels, star trackers, monitoring cameras, and antennae - The monitoring camera bracket associated non destructive analysis in this study\cite{galleguillos2023ndt}. As for NASA's Psyche mission, the asteroid orbiter spacecraft was launched on October 2023 and is on its journey to the asteroid Psyche, covering over 2.2 billion miles to arrive by 2029 \cite{nasaPsycheNASA, nasaThingsKnow}. The spacecraft asteroid orbiter integrated EBM produced AM nodes for the spacecraft Solar Electric Propulsion (SEP) chassis. The nodes interfaced with carbon composite tubes and were ideally suited for AM due to their complex geometries \cite{zenithtecnicaZenithTecnica}. 

\begin{figure}
    \centering
    \includegraphics[width=1\linewidth]{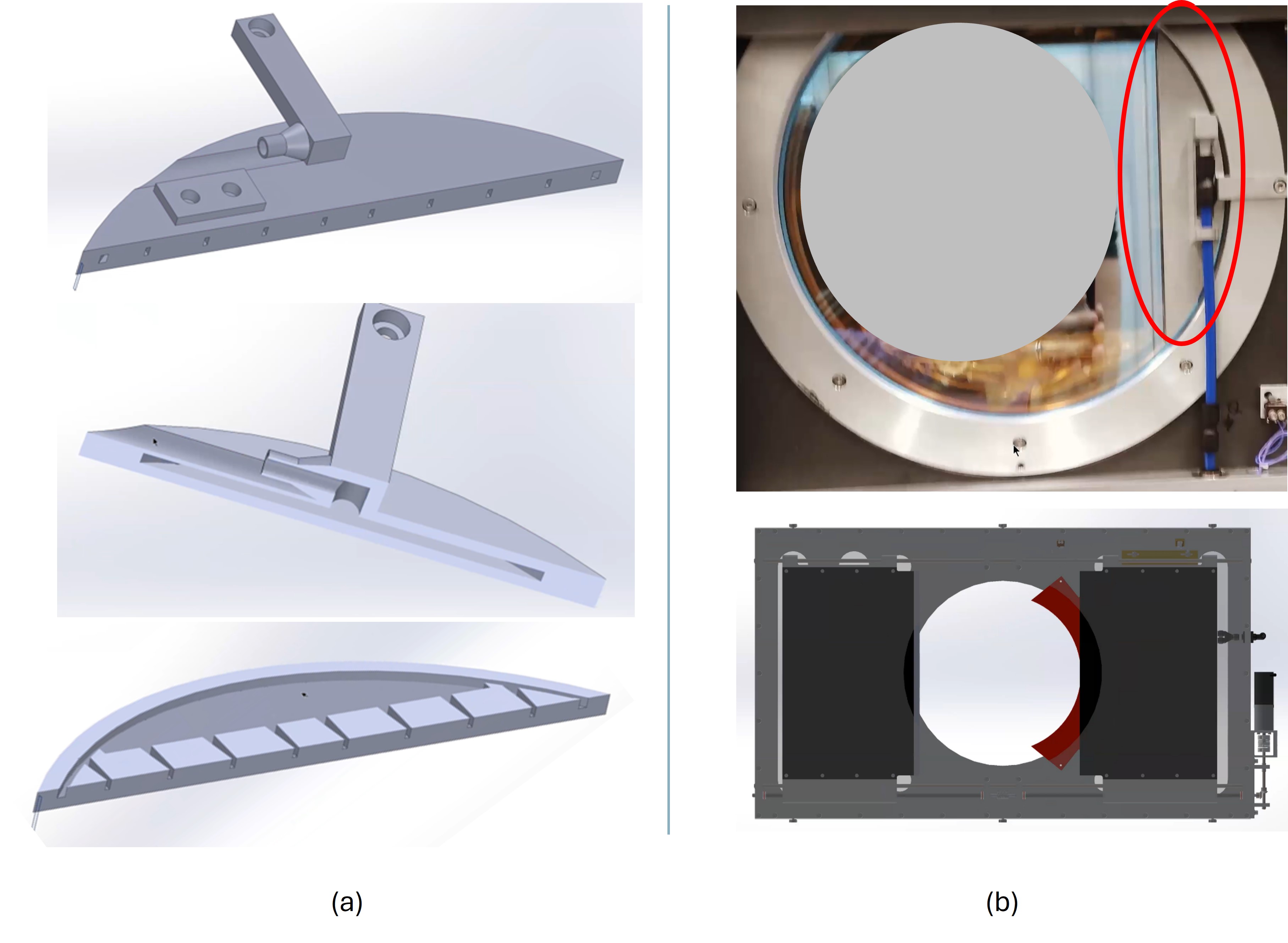}
    \caption{Window air blower used to remove condensation from the instrument window. The 3D renderings and 3D cross sections can be seen in (a). The assembled Stainless Steel 316L PBF 3D printed model can be seen integrated in GTC CIRCE instrument and circled in red (b, top). After successful use in CIRCE, two more window blowers can also be seen in GTC MIRADAS rendering in red (b,bottom). Project done by Stelter D. from UCO. Image credits: Stelter D.}
    \label{fig:denowindow}
    \end{figure}

Ground-based application of brackets include previously unreported AM components in Gran Telescopio Canarias (GTC). These include a window air blower (Figure \ref{fig:denowindow}), which directed compressed air to reduce condensation at the instrument window. The part was 3D printed using LPBF in 316L stainless steel and was suitable for AM due to its internal intricate manifold channels made custom for the 350mm instrument window. The part was implemented in GTC's Canarias InfraRed Camera Experiment (CIRCE) instrument, and upon successful use, two more were custom made and implemented for another GTC instrument called Multi-object InfraRed Absorption Spectrograph (MIRADAS). Other brackets were implemented in GTC CIRCE instrument, this time, operating under high vacuum and 3D printed in ABS. These components provided direct protection of cables (Figure \ref{fig:groundbraqcket} d) or provided means of attachment points to prevent long cables from tangling and striking important hardware during the instrument rotation (Figure \ref{fig:groundbraqcket} c). To mitigate the risk of outgassing particles from the 3D printed polymer components freezing on the detector, these components underwent vacuum baking before being installed at a considerable distance from the detector.

Another ground-based application include Multi-Object Optical and Near-infrared Spectrograph (MOONS) instrument in VLT, which also implements two polymer components produced using FDM AM process. The first one is an ASA polymer produced air duct and second one is a cable support bracket 3D printed in Antero polymer material. ASA was used for its relatively high thermal resistance and dimensional stability while Antero was used for its electro-static dissipative (ESD) functionality, making it ideal for cable management bracket applications, as previously implemented by NASA to hold fiber-optic cables, in its 2018 ICESat-2 satellite \cite{padtincPrintingSpace}. More information on AM brackets can be found in this review from literature \cite{samal20223d}.

\subsubsection{Tooling}

Tooling refers to the various tools, fixtures, and equipment used in the manufacturing and assembly of parts and products. In mechanical engineering, tooling encompasses a wide range of items, including cutting tools, molds, dies, jigs, gauges, and other devices that ensure precision and efficiency in production processes. Effective and suitable tooling is necessary for achieving high-quality, cost effective and repeatable results in manufacturing.

\begin{figure}
    \centering
    \includegraphics[width=0.65\linewidth]{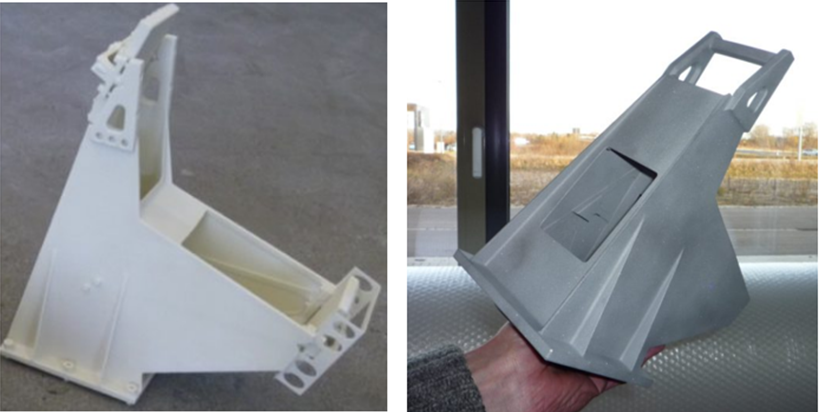}
    \caption{VLT MUSE guiding probe supporting arm 3D printed using binder jetting in PMMA polymer. The 3D printed part was used to create a wax model followed by ceramic mold (left) and eventually, the aluminium cast part (right)\cite{frank2014design}. Image credits: Frank, et al,. (2014)~\cite{frank2014design}}
    \label{fig:1sttooling}
\end{figure}

For instance, an AM tooling ground-based application has been implemented in the VLT Multi Unit Spectroscopic Explorer (MUSE) instrument. The objective was to manufacture a guiding probe supporting arm, which can be considered an optomechanical structure. To achieve this, a 3D-printed polymer tooling was created using binder jetting with polymethyl methacrylate (PMMA) material \cite{voxeljetMoldingPrinting}. The obtained model was infiltrated with wax and then used to make a ceramic mold. This ceramic mold was then used to cast aluminum (G-AlSi10Mg T6). The final part was a consolidated monolithic piece with high rigidity, saving material waste from machining alternatives and avoiding the assembly or simulation complications of using multiple parts \cite{frank2014design}. The 3D printed PMMA part and final cast can be seen in Figure \ref{fig:1sttooling}.   

In another study, AM was leveraged to produce a hyperspectral imaging spectrometer named ``Spectrolite'' for satellite applications. Unlike the previous example, this process skipped the polymer step and directly 3D printed the part in wax. The wax part was then used to create a ceramic shell by repeated dipping in a ceramic slurry, followed by a lost wax process. The material chosen for casting was aluminum alloy A357 (EN AC 42200) due to its excellent welding characteristics and strength \cite{van2016high}. 

The casting process allowed for a minimum wall thickness of 2 mm, and any flaws could be repaired using grinding or welding. The final part was within dimensional tolerance, as measured by an optical scan, which showed a deviation of ±0.55 mm. A cost comparison revealed that producing a single housing using traditional milling was 3\% cheaper. However, the hybrid AM and casting method became more cost-effective when producing multiple units, being for example 11\% cheaper for four housings and 21\% cheaper for 10. However, mirror interfaces in the cast require higher precision than the casting process could achieve, necessitating post-machining to reach a \SI{20}{\micro\meter} accuracy. AM provided significant design freedom, enabling the cost-effective production of high-grade spectrometers. These spectrometers can be deployed on nano and microsatellites for Earth observation \cite{van2016high}.

\begin{figure}
    \centering
    \includegraphics[width=0.65\linewidth]{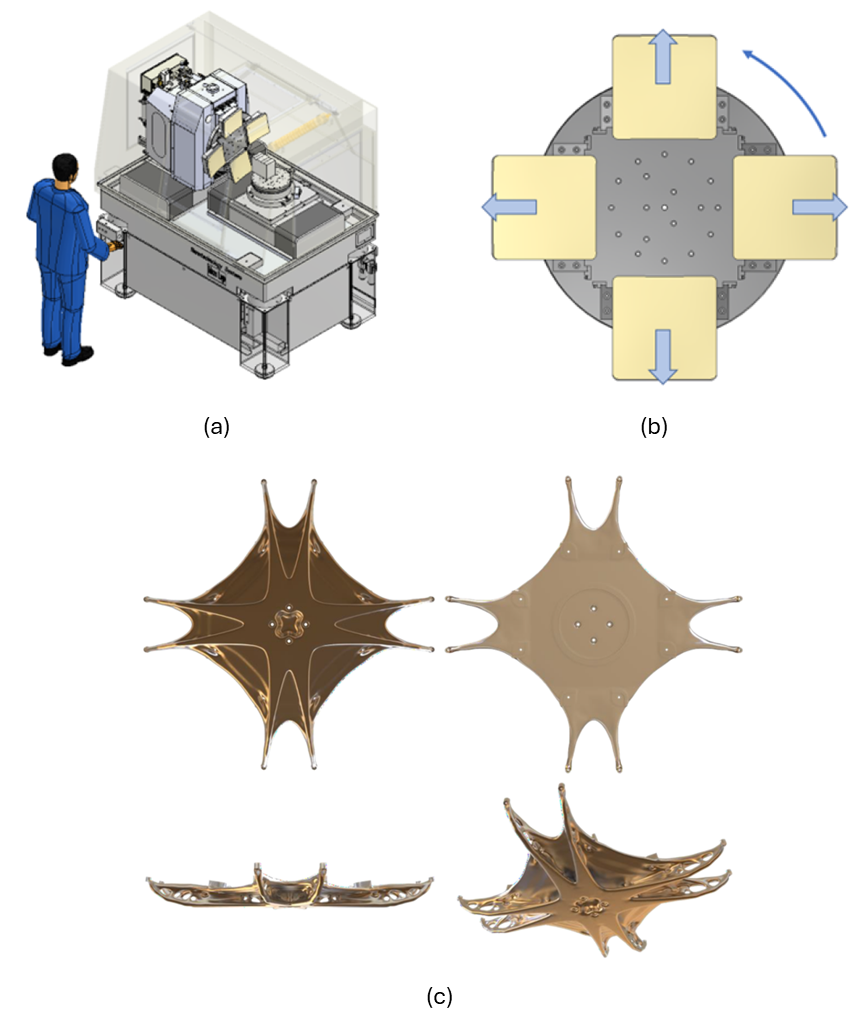}
    \caption{ Single-point DT machine (a) and example of a conventional fixture exerting rotational forces on machined mirrors (b). Topology  optimised fixture  design made for AlSi10Mg~\cite{bourgenot2024topology}. Image credits: Bourgenot, et al,. (2024)~\cite{bourgenot2024topology} }
    \label{fig:tooling3}
\end{figure}

Mirror production can also benefit from using AM tooling applications to create machining fixtures for the single-point DT process. For example, topology optimisation can be used to reduce a mirror machining fixture mass while improving its stiffness, which helps in minimising deformations from rotational and cutting forces that can affect optical performance. Using FEM, a study\cite{bourgenot2024topology} compared conventionally machined fixtures with those produced using AM. The results show that AM designs can achieve a 68\% mass reduction, enabling a single operator to handle the assembly without specialised lifting equipment. Additionally, these optimised designs significantly reduced deformation caused by rotational forces by up to 86\% and deformation caused by cutting forces by up to 51\%. The original and optimised fixture designs can be seen in Figure \ref{fig:tooling3}. Other case studies from literature include using AM to produce investment casting tooling for AlBeCast-910 mirrors\cite{sweeney2015application}, as well as different types of tooling that can be used for mirror polishing \cite{brunelle2017current}.

\section{Discussion and Conclusion}
\label{sec:discussionandconclusion}

\begin{figure}
    \centering
    \includegraphics[width=1\linewidth]{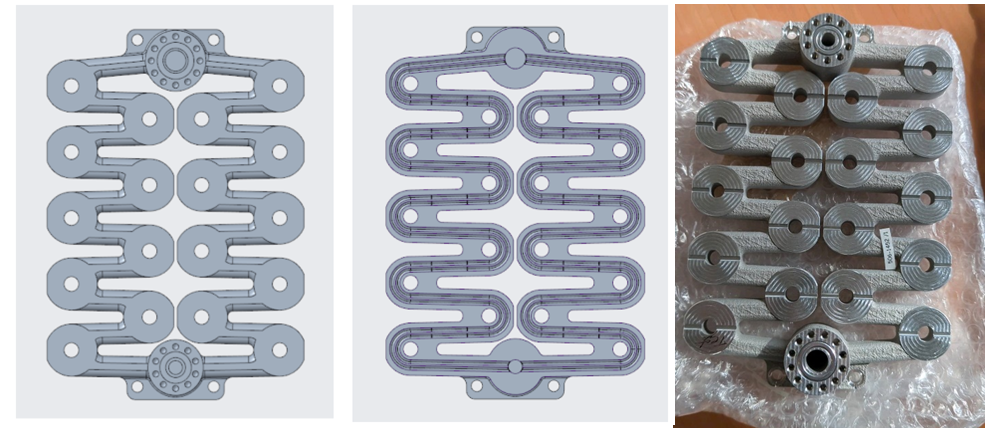}
    \caption{Heat exchanger 3D printed using LPBF process in Scalmalloy aluminium material for ELT's METIS instrument. The heat exchanger has passed leak test and is planned to be used under high vacuum. Done by Laun W. from Max-Planck Institute for Astronomy Heidelberg. Image credits: Laun W.}
    \label{fig:heatexchanger}
\end{figure}

In this review paper, an increase in AM research within astronomical instrumentation has been documented. Further analysis uncovered research trends, as well as gaps and challenges of using AM. Notably, mirrors have emerged as the most researched component. This finding suggests a likely rise in the integration of AM mirrors in upcoming projects, facilitated by future publications detailing AM quality assurance and implementation procedures. Case studies reporting implementation are crucial in the risk-averse field of astronomy, where heritage components often dominate projects, highlighting the necessity for setting precedents. A similar trend has been observed for optomechanical structures and compliant mechanisms, which are relatively less affected by the presence of AM porosity, and where the maturity and readiness of AM have been increasingly demonstrated. Especially in small to medium-sized components, topology optimised metal parts produced through AM PBF have proven to be lightweight, consolidated into fewer components, and compliant with mechanical properties as well as vibration or vacuum criteria. Again, these clear benefits suggest a need for more adoption studies as well as further research into AM for larger components using processes like DED. As for brackets, this review suggests that they are currently the most adopted AM component in the field of astronomy. Compiled studies show brackets being adopted in recently launched interplanetary spacecraft, asteroid orbiters as well as in ground-based observatories like GTC (CIRCE and MIRADAS instrument) and VLT (MOONS instrument). Although brackets are relatively less critical components, this high level of adoption clearly demonstrates an appetite in using AM, which has led to successful applications and can facilitate adoption in more vital areas of astronomical instrumentation. As for tooling, this type of application has been shown in this study to be the least researched. Compiled studies show that AM tooling applications for mirrors or optomechanical structures represent a hybrid terrain where researchers and engineers can harness the benefits of established manufacturing techniques like casting while leveraging on the benefits of AM. Another research gap lies in the field of heat exchangers, with only a single case study currently documented (see Figure \ref{fig:heatexchanger}), and implemented in ELT's METIS instrument.

In conclusion, AM is not a replacement for conventional manufacturing methods, but rather a manufacturing method that is well suited for custom parts, one-offs, and multi-loaded components, which are common attributes of astronomical hardware. However, AM adoption in astronomy is low compared to fields like aerospace and medical, which are actively using AM for end use components and in applications that are beyond prototyping. AM still has its universal sets of challenges, like quality assurance and upskilling. However, the increasing adoption and promising research outcomes shown in this study suggests that AM can play the role of an enabling technology that play a significant role in pushing the limits of astronomical instrumentation manufacturing techniques.

\acknowledgments 
 
The authors gratefully acknowledge UKRI for the funding through UKRI Future Leaders Fellowship project: Printing the future of space telescopes - MR/T042230/1.

\bibliography{report} 
\bibliographystyle{spiebib} 

\clearpage
\section{Appendices}

\subsection{A.1 List of Compiled Case Studies and Other case studies}
\label{sec:A.1}

\clearpage

\begin{table}[]
\begin{tabular}{|ccccccc|}
\hline
\multicolumn{7}{|c|}{\cellcolor{gray!20}Mirrors} \\ \hline
\multicolumn{1}{|c|}{\multirow{2}{*}{Year}} &
  \multicolumn{1}{c|}{\multirow{2}{*}{Ref.}} &
  \multicolumn{1}{c|}{\multirow{2}{*}{Environ.}} &
  \multicolumn{1}{c|}{\multirow{2}{*}{Readiness}} &
  \multicolumn{2}{c|}{Material} &
  \multirow{2}{*}{AM Process} \\ \cline{5-6}
\multicolumn{1}{|c|}{} &
  \multicolumn{1}{c|}{} &
  \multicolumn{1}{c|}{} &
  \multicolumn{1}{c|}{} &
  \multicolumn{1}{c|}{Type} &
  \multicolumn{1}{c|}{Variant} &
   \\ \hline
\multicolumn{1}{|c|}{\multirow{5}{*}{2015}} &
  \multicolumn{1}{c|}{\multirow{5}{*}{\cite{sweeney2015application}}} &
  \multicolumn{1}{c|}{Space} &
  \multicolumn{1}{c|}{Research} &
  \multicolumn{1}{c|}{Metal} &
  \multicolumn{1}{c|}{Aluminium \_AlSi10Mg} &
  EBM \\ \cline{3-7} 
\multicolumn{1}{|c|}{} &
  \multicolumn{1}{c|}{} &
  \multicolumn{1}{c|}{Space} &
  \multicolumn{1}{c|}{Research} &
  \multicolumn{1}{c|}{Metal} &
  \multicolumn{1}{c|}{Titanium \_ Ti6Al4V} &
  SLS \\ \cline{3-7} 
\multicolumn{1}{|c|}{} &
  \multicolumn{1}{c|}{} &
  \multicolumn{1}{c|}{Space} &
  \multicolumn{1}{c|}{Research} &
  \multicolumn{1}{c|}{Polymer} &
  \multicolumn{1}{c|}{PEKK Carbon reinforced} &
  Binder jetting \\ \cline{3-7} 
\multicolumn{1}{|c|}{} &
  \multicolumn{1}{c|}{} &
  \multicolumn{1}{c|}{Space} &
  \multicolumn{1}{c|}{Research} &
  \multicolumn{1}{c|}{Metal} &
  \multicolumn{1}{c|}{Stainless steel \_ 420SS + bronze} &
  Binder jetting \\ \cline{3-7} 
\multicolumn{1}{|c|}{} &
  \multicolumn{1}{c|}{} &
  \multicolumn{1}{c|}{Space} &
  \multicolumn{1}{c|}{Research} &
  \multicolumn{1}{c|}{Metal} &
  \multicolumn{1}{c|}{Inconel \_ Inconel 625} &
  EBM \\ \hline
\multicolumn{1}{|c|}{\multirow{3}{*}{2015}} &
  \multicolumn{1}{c|}{\multirow{3}{*}{\cite{herzog2015optical}}} &
  \multicolumn{1}{c|}{Space} &
  \multicolumn{1}{c|}{Research} &
  \multicolumn{1}{c|}{Metal} &
  \multicolumn{1}{c|}{Titanium \_ Ti6Al4V} &
  EBM \\ \cline{3-7} 
\multicolumn{1}{|c|}{} &
  \multicolumn{1}{c|}{} &
  \multicolumn{1}{c|}{Space} &
  \multicolumn{1}{c|}{Research} &
  \multicolumn{1}{c|}{Metal} &
  \multicolumn{1}{c|}{Aluminium \_ AlSi10Mg} &
  LPBF \\ \cline{3-7} 
\multicolumn{1}{|c|}{} &
  \multicolumn{1}{c|}{} &
  \multicolumn{1}{c|}{Space} &
  \multicolumn{1}{c|}{Research} &
  \multicolumn{1}{c|}{Metal} &
  \multicolumn{1}{c|}{Titanium \_ Ti6Al4V} &
  SLA \\ \hline
\multicolumn{1}{|c|}{\multirow{3}{*}{2017}} &
  \multicolumn{1}{c|}{\multirow{3}{*}{\cite{atkins2017additive}}} &
  \multicolumn{1}{c|}{Space} &
  \multicolumn{1}{c|}{Research} &
  \multicolumn{1}{c|}{Metal} &
  \multicolumn{1}{c|}{Aluminium \_ AlSi10Mg} &
  SLS \\ \cline{3-7} 
\multicolumn{1}{|c|}{} &
  \multicolumn{1}{c|}{} &
  \multicolumn{1}{c|}{Space} &
  \multicolumn{1}{c|}{Research} &
  \multicolumn{1}{c|}{Polymer} &
  \multicolumn{1}{c|}{Accura Bluestone} &
  LPBF \\ \cline{3-7} 
\multicolumn{1}{|c|}{} &
  \multicolumn{1}{c|}{} &
  \multicolumn{1}{c|}{Space} &
  \multicolumn{1}{c|}{Research} &
  \multicolumn{1}{c|}{Polymer} &
  \multicolumn{1}{c|}{Duraform Glass filled nylon} &
  LPBF \\ \hline
\multicolumn{1}{|c|}{2017} &
  \multicolumn{1}{c|}{\cite{feldman2017design}} &
  \multicolumn{1}{c|}{Space} &
  \multicolumn{1}{c|}{Research} &
  \multicolumn{1}{c|}{Metal} &
  \multicolumn{1}{c|}{Aluminium \_ AlSi10Mg} &
  LPBF \\ \hline
\multicolumn{1}{|c|}{2017} &
  \multicolumn{1}{c|}{\cite{woodard2017progress}} &
  \multicolumn{1}{c|}{Space} &
  \multicolumn{1}{c|}{Research} &
  \multicolumn{1}{c|}{Metal} &
  \multicolumn{1}{c|}{Aluminium \_ AlSi7Mg0.3} &
  LPBF \\ \hline
\multicolumn{1}{|c|}{\multirow{4}{*}{2017}} &
  \multicolumn{1}{c|}{\multirow{4}{*}{\cite{whitsitt2017current}}} &
  \multicolumn{1}{c|}{Space} &
  \multicolumn{1}{c|}{Research} &
  \multicolumn{1}{c|}{Metal} &
  \multicolumn{1}{c|}{Invar \_ FeNi36} &
  LPBF \\ \cline{3-7} 
\multicolumn{1}{|c|}{} &
  \multicolumn{1}{c|}{} &
  \multicolumn{1}{c|}{Space} &
  \multicolumn{1}{c|}{Research} &
  \multicolumn{1}{c|}{Metal} &
  \multicolumn{1}{c|}{Aluminium \_ AlSi10Mg} &
  SLA \\ \cline{3-7} 
\multicolumn{1}{|c|}{} &
  \multicolumn{1}{c|}{} &
  \multicolumn{1}{c|}{Space} &
  \multicolumn{1}{c|}{Research} &
  \multicolumn{1}{c|}{Metal} &
  \multicolumn{1}{c|}{Stainless steel \_ 17-4 SS} &
  LPBF \\ \cline{3-7} 
\multicolumn{1}{|c|}{} &
  \multicolumn{1}{c|}{} &
  \multicolumn{1}{c|}{Space} &
  \multicolumn{1}{c|}{Research} &
  \multicolumn{1}{c|}{Ceramic} &
  \multicolumn{1}{c|}{Alumina} &
  LPBF \\ \hline
\multicolumn{1}{|c|}{\multirow{3}{*}{2018}} &
  \multicolumn{1}{c|}{\multirow{3}{*}{\cite{atkins2018topological}}} &
  \multicolumn{1}{c|}{Space} &
  \multicolumn{1}{c|}{Research} &
  \multicolumn{1}{c|}{Metal} &
  \multicolumn{1}{c|}{Aluminium \_ AlSi10Mg} &
  LPBF \\ \cline{3-7} 
\multicolumn{1}{|c|}{} &
  \multicolumn{1}{c|}{} &
  \multicolumn{1}{c|}{Space} &
  \multicolumn{1}{c|}{Research} &
  \multicolumn{1}{c|}{Metal} &
  \multicolumn{1}{c|}{Aluminium \_ AlSi10Mg} &
  LPBF \\ \cline{3-7} 
\multicolumn{1}{|c|}{} &
  \multicolumn{1}{c|}{} &
  \multicolumn{1}{c|}{Space} &
  \multicolumn{1}{c|}{Research} &
  \multicolumn{1}{c|}{Metal} &
  \multicolumn{1}{c|}{Aluminium \_ AlSi10Mg} &
  LPBF \\ \hline
\multicolumn{1}{|c|}{\multirow{2}{*}{2019}} &
  \multicolumn{1}{c|}{\multirow{2}{*}{\cite{eberle2019additive}}} &
  \multicolumn{1}{c|}{Space} &
  \multicolumn{1}{c|}{Research} &
  \multicolumn{1}{c|}{Metal} &
  \multicolumn{1}{c|}{Aluminium \_ ALSi40} &
  LPBF \\ \cline{3-7} 
\multicolumn{1}{|c|}{} &
  \multicolumn{1}{c|}{} &
  \multicolumn{1}{c|}{Space} &
  \multicolumn{1}{c|}{Research} &
  \multicolumn{1}{c|}{Metal} &
  \multicolumn{1}{c|}{Aluminium \_ ALSi40} &
  NS \\ \hline
\multicolumn{1}{|c|}{2018} &
  \multicolumn{1}{c|}{\cite{hilpert2018precision}} &
  \multicolumn{1}{c|}{Space} &
  \multicolumn{1}{c|}{Research} &
  \multicolumn{1}{c|}{Metal} &
  \multicolumn{1}{c|}{Aluminium \_ AlSi12 + NiP} &
  NS \\ \hline
\multicolumn{1}{|c|}{2018} &
  \multicolumn{1}{c|}{\cite{mooney2018advanced}} &
  \multicolumn{1}{c|}{Space} &
  \multicolumn{1}{c|}{Research} &
  \multicolumn{1}{c|}{NS} &
  \multicolumn{1}{c|}{NS} &
  LPBF \\ \hline
\multicolumn{1}{|c|}{2019} &
  \multicolumn{1}{c|}{\cite{goodman2019ultra}} &
  \multicolumn{1}{c|}{Space} &
  \multicolumn{1}{c|}{Research} &
  \multicolumn{1}{c|}{Metal} &
  \multicolumn{1}{c|}{Ceramic \_ RoboSiC} &
  LPBF \\ \hline
\multicolumn{1}{|c|}{2019} &
  \multicolumn{1}{c|}{\cite{hilpert2019design}} &
  \multicolumn{1}{c|}{Space} &
  \multicolumn{1}{c|}{Research} &
  \multicolumn{1}{c|}{Metal} &
  \multicolumn{1}{c|}{Aluminium \_ AlSi40 + NiP} &
  LPBF \\ \hline
\multicolumn{1}{|c|}{\multirow{5}{*}{2019}} &
  \multicolumn{1}{c|}{\multirow{5}{*}{\cite{atkins2019additively}}} &
  \multicolumn{1}{c|}{Space} &
  \multicolumn{1}{c|}{Research} &
  \multicolumn{1}{c|}{Metal} &
  \multicolumn{1}{c|}{Aluminium \_ AlSi10Mg} &
  LPBF \\ \cline{3-7} 
\multicolumn{1}{|c|}{} &
  \multicolumn{1}{c|}{} &
  \multicolumn{1}{c|}{Space} &
  \multicolumn{1}{c|}{Research} &
  \multicolumn{1}{c|}{Metal} &
  \multicolumn{1}{c|}{Aluminium \_ AlSi10Mg} &
  LPBF \\ \cline{3-7} 
\multicolumn{1}{|c|}{} &
  \multicolumn{1}{c|}{} &
  \multicolumn{1}{c|}{Space} &
  \multicolumn{1}{c|}{Research} &
  \multicolumn{1}{c|}{Metal} &
  \multicolumn{1}{c|}{Aluminium \_ AlSi10Mg + NiP} &
  EBM \\ \cline{3-7} 
\multicolumn{1}{|c|}{} &
  \multicolumn{1}{c|}{} &
  \multicolumn{1}{c|}{Space} &
  \multicolumn{1}{c|}{Research} &
  \multicolumn{1}{c|}{Metal} &
  \multicolumn{1}{c|}{Aluminium \_ AlSi10Mg + NiP} &
  NS \\ \cline{3-7} 
\multicolumn{1}{|c|}{} &
  \multicolumn{1}{c|}{} &
  \multicolumn{1}{c|}{Space} &
  \multicolumn{1}{c|}{Research} &
  \multicolumn{1}{c|}{Metal} &
  \multicolumn{1}{c|}{Titanium \_ Ti6Al4V} &
  LPBF \\ \hline
\multicolumn{1}{|c|}{2019} &
  \multicolumn{1}{c|}{\cite{atkins2019lightweighting}} &
  \multicolumn{1}{c|}{Space} &
  \multicolumn{1}{c|}{Research} &
  \multicolumn{1}{c|}{} &
  \multicolumn{1}{c|}{NS \_ NS} &
  SLA \\ \hline
\multicolumn{1}{|c|}{\multirow{2}{*}{2020}} &
  \multicolumn{1}{c|}{\multirow{2}{*}{\cite{roulet2020use}}} &
  \multicolumn{1}{c|}{Space} &
  \multicolumn{1}{c|}{Research} &
  \multicolumn{1}{c|}{Metal} &
  \multicolumn{1}{c|}{Titanium \_ Ti6Al4V} &
  EBM \\ \cline{3-7} 
\multicolumn{1}{|c|}{} &
  \multicolumn{1}{c|}{} &
  \multicolumn{1}{c|}{Space} &
  \multicolumn{1}{c|}{Research} &
  \multicolumn{1}{c|}{Ceramic} &
  \multicolumn{1}{c|}{Ceramic \_ Alumina} &
  LPBF \\ \hline
\multicolumn{1}{|c|}{2020} &
  \multicolumn{1}{c|}{\cite{snell2020additive}} &
  \multicolumn{1}{c|}{Space} &
  \multicolumn{1}{c|}{Research} &
  \multicolumn{1}{c|}{Metal} &
  \multicolumn{1}{c|}{Titanium \_ Ti6Al4V} &
  NS \\ \hline
\multicolumn{1}{|c|}{2020} &
  \multicolumn{1}{c|}{\cite{tan2020design}} &
  \multicolumn{1}{c|}{Space} &
  \multicolumn{1}{c|}{Research} &
  \multicolumn{1}{c|}{Metal} &
  \multicolumn{1}{c|}{Aluminium \_ AlSi10Mg} &
  Binder Jetting \\ \hline
\multicolumn{1}{|c|}{2020} &
  \multicolumn{1}{c|}{\cite{goodman2020results}} &
  \multicolumn{1}{c|}{Space} &
  \multicolumn{1}{c|}{Research} &
  \multicolumn{1}{c|}{Ceramic} &
  \multicolumn{1}{c|}{RoboSiC} &
  NS \\ \hline
\multicolumn{1}{|c|}{2020} &
  \multicolumn{1}{c|}{\cite{horvath2020grinding}} &
  \multicolumn{1}{c|}{NS} &
  \multicolumn{1}{c|}{Research} &
  \multicolumn{1}{c|}{Ceramic} &
  \multicolumn{1}{c|}{SiC + CVI} &
  EBM \\ \hline
\multicolumn{1}{|c|}{2020} &
  \multicolumn{1}{c|}{\cite{farkas2020freeform}} &
  \multicolumn{1}{c|}{Ground} &
  \multicolumn{1}{c|}{Research} &
  \multicolumn{1}{c|}{NS} &
  \multicolumn{1}{c|}{NS} &
  Binder Jetting \\ \hline
\multicolumn{1}{|c|}{2020} &
  \multicolumn{1}{c|}{\cite{schnetler2020h2020}} &
  \multicolumn{1}{c|}{Space} &
  \multicolumn{1}{c|}{Research} &
  \multicolumn{1}{c|}{Metal} &
  \multicolumn{1}{c|}{Titanium \_ Ti6Al4V} &
  LPBF \\ \hline
\multicolumn{1}{|c|}{2021} &
  \multicolumn{1}{c|}{\cite{xu2021ultra}} &
  \multicolumn{1}{c|}{Space} &
  \multicolumn{1}{c|}{Research} &
  \multicolumn{1}{c|}{Ceramic} &
  \multicolumn{1}{c|}{SiC} &
  LPBF \\ \hline
\multicolumn{1}{|c|}{2022} &
  \multicolumn{1}{c|}{\multirow{3}{*}{\cite{paenoi2022lightweight}}} &
  \multicolumn{1}{c|}{Space} &
  \multicolumn{1}{c|}{Research} &
  \multicolumn{1}{c|}{Metal} &
  \multicolumn{1}{c|}{Aluminium \_ NS} &
  LPBF \\ \cline{1-1} \cline{3-7} 
\multicolumn{1}{|c|}{2022} &
  \multicolumn{1}{c|}{} &
  \multicolumn{1}{c|}{Space} &
  \multicolumn{1}{c|}{Research} &
  \multicolumn{1}{c|}{Metal} &
  \multicolumn{1}{c|}{Aluminium \_ NS} &
  LPBF \\ \cline{1-1} \cline{3-7} 
\multicolumn{1}{|c|}{2022} &
  \multicolumn{1}{c|}{} &
  \multicolumn{1}{c|}{Space} &
  \multicolumn{1}{c|}{Research} &
  \multicolumn{1}{c|}{Metal} &
  \multicolumn{1}{c|}{Aluminium \_ NS} &
  LPBF \\ \hline
\multicolumn{1}{|c|}{2023} &
  \multicolumn{1}{c|}{\cite{westsik2023design}} &
  \multicolumn{1}{c|}{Space} &
  \multicolumn{1}{c|}{Research} &
  \multicolumn{1}{c|}{Metal} &
  \multicolumn{1}{c|}{Aluminium \_ AlSi10Mg} &
  LPBF \\ \hline
\end{tabular}
    \caption{Compiled case studies covering AM applications for mirrors.}
    \label{tab:Compilation1}
\end{table}

\clearpage

\begin{table}[]
\begin{tabular}{|llccccc|}
\hline
\multicolumn{7}{|c|}{\cellcolor{gray!20}Optomechanical Structures} \\ \hline
\multicolumn{1}{|c|}{\multirow{2}{*}{Year}} &
  \multicolumn{1}{c|}{\multirow{2}{*}{Ref.}} &
  \multicolumn{1}{c|}{\multirow{2}{*}{Environ.}} &
  \multicolumn{1}{c|}{\multirow{2}{*}{Readiness}} &
  \multicolumn{2}{c|}{Material} &
  \multirow{2}{*}{AM Process} \\ \cline{5-6}
\multicolumn{1}{|c|}{} &
  \multicolumn{1}{c|}{} &
  \multicolumn{1}{c|}{} &
  \multicolumn{1}{c|}{} &
  \multicolumn{1}{c|}{Type} &
  \multicolumn{1}{c|}{Variant} &
  \multicolumn{1}{c|}{} \\ \hline
\multicolumn{1}{|l|}{2018} &
  \multicolumn{1}{l|}{\cite{heidler2018additive}} &
  \multicolumn{1}{c|}{Space} &
  \multicolumn{1}{c|}{Research} &
  \multicolumn{1}{c|}{Metal} &
  \multicolumn{1}{c|}{Aluminium\_ALSi40} &
  \multicolumn{1}{c|}{LPBF} \\ \hline
\multicolumn{1}{|l|}{2018} &
  \multicolumn{1}{l|}{\cite{steele20183d}} &
  \multicolumn{1}{c|}{Ground} &
  \multicolumn{1}{c|}{Research} &
  \multicolumn{1}{c|}{Polymer} &
  \multicolumn{1}{c|}{ABS} &
  \multicolumn{1}{c|}{FDM} \\ \hline
\multicolumn{1}{|l|}{2018} &
  \multicolumn{1}{l|}{\cite{schimmerohn2018additive}} &
  \multicolumn{1}{c|}{Space} &
  \multicolumn{1}{c|}{Research} &
  \multicolumn{1}{c|}{Metal} &
  \multicolumn{1}{c|}{Aluminium\_AlSi10Mg} &
  \multicolumn{1}{c|}{LPBF} \\ \hline
\multicolumn{1}{|l|}{2019} &
  \multicolumn{1}{l|}{\cite{heidler2019topology}} &
  \multicolumn{1}{c|}{Space} &
  \multicolumn{1}{c|}{Research} &
  \multicolumn{1}{c|}{Metal} &
  \multicolumn{1}{c|}{Aluminium\_AlSi40} &
  \multicolumn{1}{c|}{LPBF} \\ \hline
\multicolumn{1}{|l|}{2020} &
  \multicolumn{1}{l|}{\cite{goodman2020results}} &
  \multicolumn{1}{c|}{Space} &
  \multicolumn{1}{c|}{Research} &
  \multicolumn{1}{c|}{Ceramic} &
  \multicolumn{1}{c|}{RoboSiC} &
  \multicolumn{1}{c|}{*} \\ \hline
\multicolumn{1}{|l|}{2020} &
  \multicolumn{1}{l|}{\cite{nessancleary3DprintedTelescope,padtincPrintingSpace}} &
  \multicolumn{1}{c|}{Space} &
  \multicolumn{1}{c|}{Implemented} &
  \multicolumn{1}{c|}{Polymer} &
  \multicolumn{1}{c|}{Ultem 9085} &
  \multicolumn{1}{c|}{FDM} \\ \hline
\multicolumn{1}{|l|}{2020} &
  \multicolumn{1}{l|}{\cite{esaPrintingTelescope}} &
  \multicolumn{1}{c|}{Space} &
  \multicolumn{1}{c|}{Research} &
  \multicolumn{1}{c|}{Metal} &
  \multicolumn{1}{c|}{Aluminium\_NS} &
  \multicolumn{1}{c|}{*} \\ \hline
\multicolumn{1}{|l|}{2021} &
  \multicolumn{1}{l|}{\cite{von2021optimization}} &
  \multicolumn{1}{c|}{Space} &
  \multicolumn{1}{c|}{Research} &
  \multicolumn{1}{c|}{Metal} &
  \multicolumn{1}{c|}{Aluminium\_AlSi40} &
  \multicolumn{1}{c|}{LPBF} \\ \hline
\multicolumn{1}{|l|}{2021} &
  \multicolumn{1}{l|}{\cite{atkins2022opticon}} &
  \multicolumn{1}{c|}{Space} &
  \multicolumn{1}{c|}{Research} &
  \multicolumn{1}{c|}{Metal} &
  \multicolumn{1}{c|}{Aluminium\_AlSi10Mg} &
  \multicolumn{1}{c|}{LPBF} \\ \hline
\multicolumn{1}{|l|}{2022} &
  \multicolumn{1}{l|}{\cite{frasch2022optical}} &
  \multicolumn{1}{c|}{Space} &
  \multicolumn{1}{c|}{Research} &
  \multicolumn{1}{c|}{Metal} &
  \multicolumn{1}{c|}{Aluminium\_AlSi10Mg} &
  \multicolumn{1}{c|}{LPBF} \\ \hline
\multicolumn{1}{|c|}{2022} &
  \multicolumn{1}{l|}{Fig.\ref{fig:spectrographhousing}} &
  \multicolumn{1}{c|}{Ground} &
  \multicolumn{1}{c|}{Research} &
  \multicolumn{1}{c|}{Polymer} &
  \multicolumn{1}{c|}{PA12} &
  \multicolumn{1}{c|}{SLS} \\ \hline
\multicolumn{1}{|l|}{2022} &
  \multicolumn{1}{l|}{\cite{pfuhl2022fully}} &
  \multicolumn{1}{c|}{*} &
  \multicolumn{1}{c|}{Research} &
  \multicolumn{1}{c|}{Polymer} &
  \multicolumn{1}{c|}{*} &
  \multicolumn{1}{c|}{Binder jetting} \\ \hline
\multicolumn{1}{|l|}{2022} &
  \multicolumn{1}{l|}{Fig.\ref{fig:Nasmyth}} &
  \multicolumn{1}{c|}{Ground} &
  \multicolumn{1}{c|}{Implemented} &
  \multicolumn{1}{c|}{Polymer} &
  \multicolumn{1}{c|}{ONYX} &
  \multicolumn{1}{c|}{FDM} \\ \hline
\multicolumn{1}{|l|}{2023} &
  \multicolumn{1}{l|}{\cite{wells2023lightweighting}} &
  \multicolumn{1}{c|}{Ground} &
  \multicolumn{1}{c|}{Research} &
  \multicolumn{1}{c|}{Metal} &
  \multicolumn{1}{c|}{Aluminium\_5356} &
  \multicolumn{1}{c|}{DED} \\ \hline
\multicolumn{1}{|l|}{2023} &
  \multicolumn{1}{l|}{\cite{schneider2023additive}} &
  \multicolumn{1}{c|}{Space} &
  \multicolumn{1}{c|}{Research} &
  \multicolumn{1}{c|}{Metal} &
  \multicolumn{1}{c|}{Titanium\_Ti6Al4V} &
  \multicolumn{1}{c|}{DED} \\ \hline
\multicolumn{1}{|l|}{2023} &
  \multicolumn{1}{l|}{\cite{fu2024lightweight}} &
  \multicolumn{1}{c|}{Space} &
  \multicolumn{1}{c|}{Research} &
  \multicolumn{1}{c|}{Metal} &
  \multicolumn{1}{c|}{Aluminium\_AlSi10Mg} &
  \multicolumn{1}{c|}{LPBF} \\ \hline
\multicolumn{7}{|c|}{\cellcolor{gray!20}Compliant Mechanisms} \\ \hline
\multicolumn{1}{|l|}{2013} &
  \multicolumn{1}{l|}{\cite{merriam2013monolithic}} &
  \multicolumn{1}{c|}{Space} &
  \multicolumn{1}{c|}{Research} &
  \multicolumn{1}{c|}{Metal} &
  \multicolumn{1}{c|}{Titanium\_Ti6Al4V} &
  \multicolumn{1}{c|}{EBM} \\ \hline
\multicolumn{1}{|l|}{2018} &
  \multicolumn{1}{l|}{\multirow{3}{*}{\cite{saudan2018compliant}}} &
  \multicolumn{1}{c|}{Space} &
  \multicolumn{1}{c|}{Research} &
  \multicolumn{1}{c|}{Metal} &
  \multicolumn{1}{c|}{Stainless steel \_ 316L} &
  \multicolumn{1}{c|}{LPBF} \\ \cline{1-1} \cline{3-7} 
\multicolumn{1}{|l|}{2018} &
  \multicolumn{1}{l|}{} &
  \multicolumn{1}{c|}{Space} &
  \multicolumn{1}{c|}{Research} &
  \multicolumn{1}{c|}{Metal} &
  \multicolumn{1}{c|}{Stainless steel \_ 17-4PH} &
  \multicolumn{1}{c|}{LPBF} \\ \cline{1-1} \cline{3-7} 
\multicolumn{1}{|l|}{2018} &
  \multicolumn{1}{l|}{} &
  \multicolumn{1}{c|}{Space} &
  \multicolumn{1}{c|}{Research} &
  \multicolumn{1}{c|}{Metal} &
  \multicolumn{1}{c|}{Stainless steel \_ CL92PH} &
  \multicolumn{1}{c|}{LPBF} \\ \hline
\multicolumn{1}{|l|}{2020} &
  \multicolumn{1}{l|}{\cite{thetpraphi2020advanced}} &
  \multicolumn{1}{c|}{Ground} &
  \multicolumn{1}{c|}{Research} &
  \multicolumn{1}{c|}{Polymer} &
  \multicolumn{1}{c|}{Terpolymer} &
  \multicolumn{1}{c|}{FDM} \\ \hline
\multicolumn{1}{|l|}{2020} &
  \multicolumn{1}{l|}{\cite{cosandierdeveloppement}} &
  \multicolumn{1}{c|}{Ground} &
  \multicolumn{1}{c|}{Research} &
  \multicolumn{1}{c|}{Metal} &
  \multicolumn{1}{c|}{Stainless steel \_ 17-4PH} &
  \multicolumn{1}{c|}{LPBF} \\ \hline
\multicolumn{1}{|l|}{2020} &
  \multicolumn{1}{l|}{\cite{rouvinet2020pulsar}} &
  \multicolumn{1}{c|}{Ground} &
  \multicolumn{1}{c|}{Research} &
  \multicolumn{1}{c|}{Metal} &
  \multicolumn{1}{c|}{Stainless steel \_ 17-4PH} &
  \multicolumn{1}{c|}{LPBF} \\ \hline
\multicolumn{1}{|c|}{\multirow{3}{*}{2022}} &
  \multicolumn{1}{l|}{\multirow{3}{*}{\cite{morris2022additively}}} &
  \multicolumn{1}{c|}{Ground} &
  \multicolumn{1}{c|}{Research} &
  \multicolumn{1}{c|}{Metal} &
  \multicolumn{1}{c|}{Aluminium\_AlSi10Mg} &
  \multicolumn{1}{c|}{LPBF} \\ \cline{3-7} 
\multicolumn{1}{|c|}{} &
  \multicolumn{1}{l|}{} &
  \multicolumn{1}{c|}{Ground} &
  \multicolumn{1}{c|}{Research} &
  \multicolumn{1}{c|}{Metal} &
  \multicolumn{1}{c|}{Stainless steel \_ 316L} &
  \multicolumn{1}{c|}{Binder jetting} \\ \cline{3-7} 
\multicolumn{1}{|c|}{} &
  \multicolumn{1}{l|}{} &
  \multicolumn{1}{c|}{Ground} &
  \multicolumn{1}{c|}{Research} &
  \multicolumn{1}{c|}{Metal} &
  \multicolumn{1}{c|}{Titanium\_Ti6Al4V} &
  \multicolumn{1}{c|}{LPBF} \\ \hline
\multicolumn{1}{|l|}{2022} &
  \multicolumn{1}{l|}{\cite{kiener2022validation}} &
  \multicolumn{1}{c|}{Space} &
  \multicolumn{1}{c|}{Research} &
  \multicolumn{1}{c|}{Metal} &
  \multicolumn{1}{c|}{Stainless steel \_ 17-4PH} &
  \multicolumn{1}{c|}{LPBF} \\ \hline
\multicolumn{7}{|c|}{\cellcolor{gray!20}Brackets} \\ \hline
\multicolumn{1}{|l|}{2016} &
  \multicolumn{1}{l|}{Fig.\ref{fig:denowindow}} &
  \multicolumn{1}{c|}{Ground} &
  \multicolumn{1}{c|}{Implemented} &
  \multicolumn{1}{c|}{Metal} &
  \multicolumn{1}{c|}{Stainless steel \_ 316L} &
  \multicolumn{1}{c|}{LPBF} \\ \hline
\multicolumn{1}{|l|}{2016} &
  \multicolumn{1}{l|}{Fig.\ref{fig:denowindow}x2} &
  \multicolumn{1}{c|}{Ground} &
  \multicolumn{1}{c|}{Implemented} &
  \multicolumn{1}{c|}{Metal} &
  \multicolumn{1}{c|}{Stainless steel \_ 316L} &
  \multicolumn{1}{c|}{LPBF} \\ \hline
\multicolumn{1}{|l|}{2016} &
  \multicolumn{1}{l|}{Fig.\ref{fig:groundbraqcket}c} &
  \multicolumn{1}{c|}{Ground} &
  \multicolumn{1}{c|}{Implemented} &
  \multicolumn{1}{c|}{Polymer} &
  \multicolumn{1}{c|}{ABS} &
  \multicolumn{1}{c|}{FDM} \\ \hline
\multicolumn{1}{|l|}{2016} &
  \multicolumn{1}{l|}{Fig.\ref{fig:groundbraqcket}d} &
  \multicolumn{1}{c|}{Ground} &
  \multicolumn{1}{c|}{Implemented} &
  \multicolumn{1}{c|}{Polymer} &
  \multicolumn{1}{c|}{ABS} &
  \multicolumn{1}{c|}{FDM} \\ \hline
\multicolumn{1}{|l|}{2021} &
  \multicolumn{1}{l|}{\cite{manil2021structural}} &
  \multicolumn{1}{c|}{Space} &
  \multicolumn{1}{c|}{Research} &
  \multicolumn{1}{c|}{Metal} &
  \multicolumn{1}{c|}{Titanium\_Ti6Al4V} &
  \multicolumn{1}{c|}{LPBF} \\ \hline
\multicolumn{1}{|l|}{\multirow{2}{*}{2022}} &
  \multicolumn{1}{l|}{\multirow{2}{*}{\cite{mcclelland2022generative}}} &
  \multicolumn{1}{c|}{\multirow{2}{*}{Space}} &
  \multicolumn{1}{c|}{\multirow{2}{*}{Research}} &
  \multicolumn{1}{c|}{\multirow{2}{*}{Metal}} &
  \multicolumn{1}{c|}{Aluminium\_AlSi10Mg} &
  \multicolumn{1}{c|}{LPBF} \\ \cline{6-7} 
\multicolumn{1}{|l|}{} &
  \multicolumn{1}{l|}{} &
  \multicolumn{1}{c|}{} &
  \multicolumn{1}{c|}{} &
  \multicolumn{1}{c|}{} &
  \multicolumn{1}{c|}{Titanium\_Ti6Al4V} &
  \multicolumn{1}{c|}{LPBF} \\ \hline
\multicolumn{1}{|l|}{2023} &
  \multicolumn{1}{l|}{\cite{voxelmattersJUICEPrinted,esaJuiceFactsheet}x13} &
  \multicolumn{1}{c|}{Space} &
  \multicolumn{1}{c|}{Implemented} &
  \multicolumn{1}{c|}{Metal} &
  \multicolumn{1}{c|}{Aluminium\_Scalmalloy} &
  \multicolumn{1}{c|}{LPBF} \\ \hline
\multicolumn{1}{|l|}{2023} &
  \multicolumn{1}{l|}{\cite{zenithtecnicaZenithTecnica}} &
  \multicolumn{1}{c|}{Space} &
  \multicolumn{1}{c|}{Implemented} &
  \multicolumn{1}{c|}{Metal} &
  \multicolumn{1}{c|}{Titanium\_Ti6Al4V} &
  \multicolumn{1}{c|}{EBM} \\ \hline
\multicolumn{1}{|l|}{2024} &
  \multicolumn{1}{l|}{Fig.\ref{fig:groundbraqcket}ax2} &
  \multicolumn{1}{c|}{Ground} &
  \multicolumn{1}{c|}{Implemented} &
  \multicolumn{1}{c|}{Polymer} &
  \multicolumn{1}{c|}{Antero} &
  \multicolumn{1}{c|}{FDM} \\ \hline
\multicolumn{1}{|l|}{2024} &
  \multicolumn{1}{l|}{Fig.\ref{fig:groundbraqcket}b} &
  \multicolumn{1}{c|}{Ground} &
  \multicolumn{1}{c|}{Implemented} &
  \multicolumn{1}{c|}{Polymer} &
  \multicolumn{1}{c|}{ASA} &
  \multicolumn{1}{c|}{FDM} \\ \hline
\multicolumn{7}{|c|}{\cellcolor{gray!20}Tooling} \\ \hline
\multicolumn{1}{|l|}{2014} &
  \multicolumn{1}{l|}{\cite{frank2014design}} &
  \multicolumn{1}{c|}{Ground} &
  \multicolumn{1}{c|}{Implemented} &
  \multicolumn{1}{c|}{Polymer} &
  \multicolumn{1}{c|}{PMMA} &
  \multicolumn{1}{c|}{Binder jetting} \\ \hline
\multicolumn{1}{|l|}{2015} &
  \multicolumn{1}{l|}{\cite{sweeney2015application}} &
  \multicolumn{1}{c|}{Space} &
  \multicolumn{1}{c|}{Research} &
  \multicolumn{1}{c|}{Polymer} &
  \multicolumn{1}{c|}{*} &
  \multicolumn{1}{c|}{SLA} \\ \hline
\multicolumn{1}{|l|}{2016} &
  \multicolumn{1}{l|}{\cite{van2016high}} &
  \multicolumn{1}{c|}{Space} &
  \multicolumn{1}{c|}{Research} &
  \multicolumn{1}{c|}{Wax} &
  \multicolumn{1}{c|}{*} &
  \multicolumn{1}{c|}{*} \\ \hline
\multicolumn{1}{|l|}{\multirow{2}{*}{2017}} &
  \multicolumn{1}{l|}{\multirow{2}{*}{\cite{brunelle2017current}}} &
  \multicolumn{1}{c|}{Space} &
  \multicolumn{1}{c|}{Implemented} &
  \multicolumn{1}{c|}{Polymer} &
  \multicolumn{1}{c|}{*} &
  \multicolumn{1}{c|}{FDM} \\ \cline{3-7} 
\multicolumn{1}{|l|}{} &
  \multicolumn{1}{l|}{} &
  \multicolumn{1}{c|}{Space} &
  \multicolumn{1}{c|}{Implemented} &
  \multicolumn{1}{c|}{Polymer} &
  \multicolumn{1}{c|}{*} &
  \multicolumn{1}{c|}{SLA} \\ \hline
\multicolumn{1}{|l|}{2024} &
  \multicolumn{1}{l|}{\cite{bourgenot2024topology}} &
  \multicolumn{1}{c|}{Space} &
  \multicolumn{1}{c|}{Research} &
  \multicolumn{1}{c|}{Aluminium} &
  \multicolumn{1}{c|}{AlSi10Mg} &
  \multicolumn{1}{c|}{*} \\ \hline
\end{tabular}
\caption{Compiled case studies covering AM applications. * refers to ``non specified''. Case studies that have multiple components are marked by a ``x'' sign, followed by the number of parts.}
\label{tab:Compilation2}
\end{table}

\clearpage

\begin{figure}
    \centering
    \includegraphics[width=0.8\linewidth]{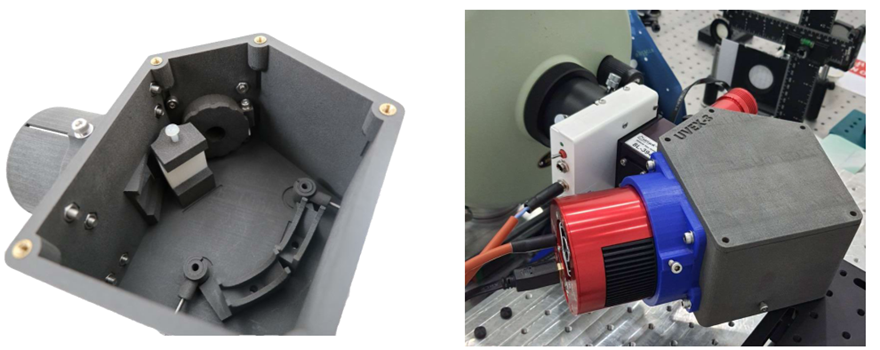}
    \caption{Spectrograph housing 3D printed using SLS AM process in PA12 polymer material. Project done Panpang P. Kanjanasakul C. and Supasri T. from National Astronomical Research Institute of Thailand (NARIT). Image credits: NARIT.}
    \label{fig:spectrographhousing}
\end{figure}
\begin{figure}
    \centering
    \includegraphics[width=1\linewidth]{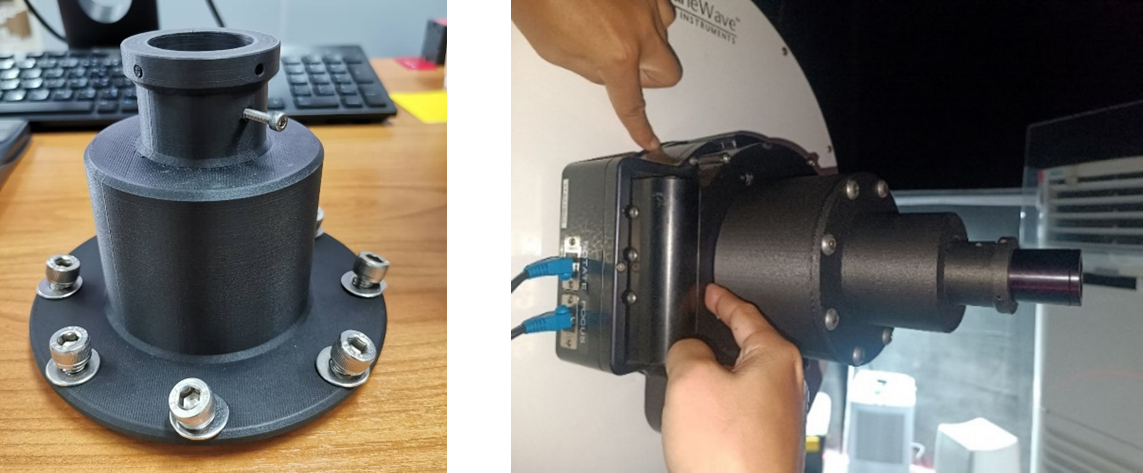}
    \caption{Nasmyth spacer 3D printed using FDM in ONYX materia. Project done by Muangnin N., Chartsiriwattana P., Prasit P. and Saladtook T. from NARIT. Image credits: NARIT.}
    \label{fig:Nasmyth}
\end{figure}

\begin{figure}
    \centering
    \includegraphics[width=1\linewidth]{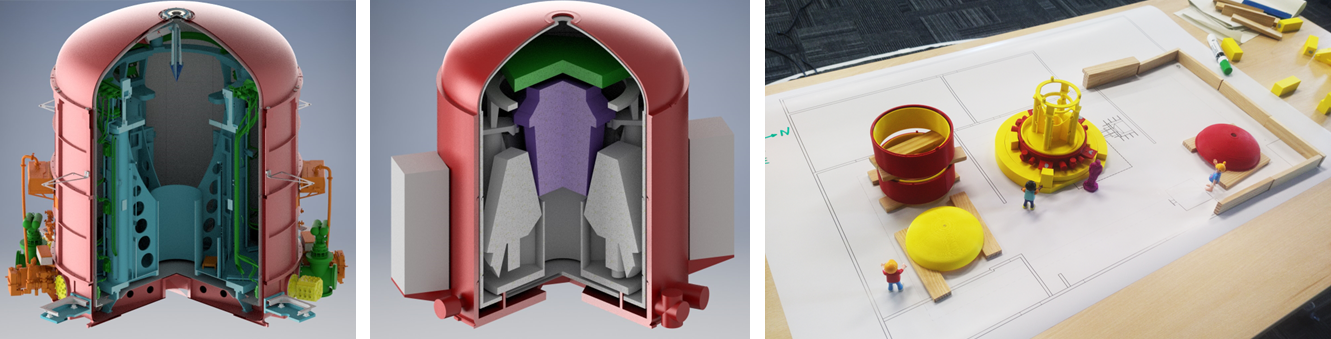}
    \caption{Orginial CAD (left) and redesign (middle) for assembly sequence planning (right). The component is an Integral Field Spectrograph (IFS) part of ELT's High Angular Resolution Monolithic Optical and Near-infrared Integral field spectrograph (HARMONI) instrument. Done by Harman J. from UKATC. Image credits: Harman J.}
    \label{fig:enter-label}
\end{figure}

\clearpage

\end{document}